\documentclass[12pt,letterpaper]{article}

\usepackage{comment}
\usepackage{pgfplots}

\pgfplotsset{
    compat=1.18,
    /pgf/number format/read comma as period
}

\usepackage{subcaption}

\usepackage{amsthm}
\usepackage{setspace}
\usepackage[multiple]{footmisc}
\usepackage{amsfonts,amssymb,amsmath}
\usepackage[margin=1.25 in]{geometry}
\usepackage{caption}
\usepackage{wasysym}
\usepackage{units}
\usepackage[multiple]{footmisc}
\usepackage[authordate,autocite=inline,backend=biber,sorting=nyt,natbib=true,useprefix=false]{biblatex-chicago}

\usepackage[scr=esstix]{mathalpha} %esstix,boondox
\usepackage[multiple]{footmisc}

\newtheorem{theorem}{Theorem}

\newtheorem{assumption}{Assumption}

\newtheorem{definition}{Definition}
\newtheorem{example}{Example}

\newtheorem{lemma}{Lemma}

\newtheorem{proposition}{Proposition}
\newtheorem{remark}{Remark}

\title{Renegotiation-Proof Cheap Talk\thanks{We are grateful for comments from participants at the Econometric Society CSW Asia Meeting 2026, the Second Paris Workshop on Games, Decisions, and Language, the Canadian Economic Theory Conference 2025, the 5th Durham Economic Theory Conference, the 2025 Conference of the Society for the Advancement of Economic Theory, the XIX GRASS Workshop, and the 2026 Economic Theory Workshop at NYU Shanghai as well as seminars at the Berlin School of Economics, University of Cyprus, Concordia University, Carleton University, Universitat de Barcelona, University of Rome Tor Vergata, and Ko\c{c} University.  Our paper has benefited from discussions with Emiliano Catonini, Michael Greinecker, Nenad Kos, Ming Li, and George Mailath. Christoph Kuzmics gratefully acknowledges financial support from the Jubiläumsfonds of the Österreichische Nationalbank (Project 18899).}}
\author{Steven Kivinen\thanks{Dept. of Economics, University of Graz, Graz, Austria.  \emph{Email:} steven.kivinen@uni-graz.at.}\and Christoph Kuzmics\thanks{Dept. of Economics, University of Graz, Graz, Austria.  \emph{Email:} christoph.kuzmics@uni-graz.at.}  }
\date{\today\\ }

\begin{document}

\maketitle

\vspace{-10pt}

\begin{abstract} 
	
\noindent An informed Expert and an uninformed Decision-Maker, with conflicting interests, engage in repeated cheap talk communication in a sequence of new decision problems. While the Decision-Maker's optimal payoff is attainable in some subgame perfect equilibrium, no payoff profile close to the Decision-Maker's optimal one is immune to renegotiation. Pareto efficient renegotiation-proof equilibria entail a compromise between the Expert and the Decision-Maker. This could involve the Expert being truthful and the Decision-Maker not fully utilizing this information to their advantage. The Decision-Maker's maximal renegotiation-proof equilibrium payoff depends on the Expert's preferences and the prior distribution of the state. In the \citet{crawford82} environment a more biased Expert harms the Decision-Maker.  In a model with transparent Expert motives a more volatile environment harms the Decision-Maker while helping the Expert.

\vspace{10pt}

	\clearpage
\end{abstract}

\section{Introduction}

In many situations, those with the authority to take an action are not those with the most precise information.  Thus, efficiency often requires effective communication. However, communication is often hindered by conflicting interests; experts have an incentive to give advice that serves their own ends. Think of bureaucrats advising their ministers, local governments soliciting funding for public projects from higher levels of government, or salespeople trying to persuade a potential consumer to buy a product. There is a large literature of cheap talk Sender-Receiver games dealing with the strategic communication that arises in such situations, starting with the seminal work by \citet{crawford82} (see \citet{sobel2013giving} for a survey). The main insights of this literature are that, when there is a conflict of interest, in equilibrium full information transmission is typically not possible. Partial information transmission, however, is possible, and the extent to which information can be transmitted depends on the details of the incentive structure. 

While this literature studies short-lived (one-shot) interactions, in many cases the Sender-Receiver relationship is longer lasting and involves a variety of ever different decision problems. We are, thus, interested in repeated interaction in settings where the Receiver has no alternative source of information: the Sender is really the only expert on the topics of interest to the Receiver. The prime example we have in mind is the long-lived relationship between a long-serving government official and an elected politician. Think of the television show ``Yes Minister'' for instance. The politician cannot fire the official; the politician also does not want to fire the official, as the official is the only one who has all the pertinent information. But they typically do not fully agree on the most desired course of action and the politician has to live with this conflict. We here mostly take the position of the politician, that is, the Receiver and decision-maker. We are particularly interested in how well the Receiver can do in such settings. 

In such repeated games, if the history of play is sufficiently public so that the truth eventually comes out, then one might expect a version of the folk theorem to apply: that almost any outcome can be sustained in a subgame perfect equilibrium (SPE) of the repeated game.\footnote{See e.g., \citet{fudenberg86} and \citet[Chapter 3]{mailath06}.} Among these, in particular, is the strategy captured by the well-known idiom, ``Fool me once, shame on you.  Fool me twice, shame on me.''\footnote{The German version is, ``Wer einmal l\"{u}gt, dem glaubt man nicht, und wenn er auch die Wahrheit spricht,'' which roughly translates to, ``If someone lies once, no one will believe them, even if they tell the truth.'' There is an older Latin idiom behind these as well.} The implicitly suggested strategy seems to be a grim trigger strategy, in which the Sender is truthful and upon any deviation detected by the Receiver, play reverts to a babbling equilibrium in which the Receiver completely ignores whatever the Sender says. Call this the F1F2 strategy. 

Any parent of a sometimes non-fully truthful child (with the parent possibly using these idioms to try to mold their child into a better human being) knows that they don't really want to go through with the threat made explicit in the F1F2 strategy. Yes, the babbling equilibrium is a Nash equilibrium and, thus, the F1F2 strategy is subgame perfect, but it is reasonable to expect a parent to strongly prefer to renegotiate with their child to avoid the babbling equilibrium that can only be bad for both. One would imagine that similar concerns would arise in any longer-lasting Sender-Receiver relationship. Indeed, concerns about renegotiation are natural in cheap talk environments, as parties can freely communicate.

To this end, in this paper the two players can renegotiate their initially agreed upon strategies, and so our notion of credibility (of threats) is stronger than subgame perfection. We consider weakly renegotiation-proof (WRP) equilibria, as defined by \citet{maskin89}, in infinitely repeated cheap talk games.  

Like the grim trigger strategy in repeated prisoners' dilemma games, F1F2 is vulnerable to renegotiation. Instead of carrying out the punishment, both players would rather go back to being cooperative, which would entail the Sender being truthful. In the repeated prisoners' dilemma the cooperative outcome can, nevertheless, be sustained by a WRP strategy profile (see e.g., \citet{vandamme89}). Can, likewise, the Receiver-optimal SPE payoff be sustained by a WRP strategy profile? 

We establish as our main result (Theorem \ref{thm:main}) that in repeated cheap-talk games any WRP equilibrium payoff is bounded away from the Receiver-optimal SPE payoff. These results hold under relatively weak assumptions about the stage game: We require that the Receiver has a unique and distinct optimal action for each state, and that the Sender and the Receiver have at least a minimal conflict of interest.\footnote{We also require that the message space is sufficiently rich, so that agents' communication is not constrained by a lack of vocabulary.}  Our result relies on Lemma 4, which extends \citet{maskin89}'s characterization of WRP equilibrium payoffs to non-finite games.

The Receiver, thus, cannot expect to obtain their maximal SPE payoff in a WRP equilibrium.\footnote{We will sometimes abuse terminology somewhat by not distinguishing between the maximum and supremum payoffs, even though this distinction matters in the formal results (which are written precisely).} Moreover, the Receiver's maximal WRP payoff (on the Pareto frontier), unlike the Receiver's maximal SPE payoff, depends on the Sender's preferences and on the prior distribution of the state. One strategy profile that always implements the Receiver maximal WRP payoff (in any repeated cheap talk game and if it exists) has, on the equilibrium path of play, the Sender being truthful and the Receiver playing a compromise action in each state that maximizes (state by state) some weighted sum of the Sender and Receiver utility. There are typically also other ways to implement the Receiver's maximal WRP payoff. The Sender can for instance, again on the equilibrium path of play, distort the truth somewhat, with the Receiver ``pretending'' to believe them, with the same ultimate resulting compromise actions being played.

We also identify a useful result (Lemma 5) that allows us to study WRP payoffs (especially on the Pareto frontier) in greater depth in different applications. This result is informative about the possible on path equilibrium play, but is about the ``cheapest'' way for the Receiver to punish any Sender deviations from being truthful. To get there, we establish a series of useful lemmas of some independent interest. These ultimately help us to identify the range of payoff profiles on the Pareto frontier that can be sustained by a WRP equilibrium in a variety of applications. 

First, we establish (in Lemma \ref{lem:revelation}) that any feasible stage-game payoff profile can also be achieved through the Sender being truthful and the Receiver appropriately adapting their strategy.\footnote{With some license, this result may be viewed as a revelation principle. Unlike the revelation principle of mechanism design, truthful reporting is not without loss in the underlying cheap-talk stage game, since the resulting change in the Receiver strategy alters the Sender's incentives. The Sender's incentives, however, can be restored through the repeated-game structure.} This also implies that any payoff profile on the Pareto frontier can be established with the Sender being truthful. For games of the \citet{crawford82} variety (with strictly convex preferences) any payoff profile on the Pareto frontier, in fact, requires the Sender to be truthful (almost surely).\footnote{Any Sender strategy that is a bijection would work as well, of course. A bijection that is not the identity function could be interpreted as being truthful just using a different language / vocabulary.} 

Second (Lemma \ref{lem:convexF}), for cheap-talk games the set of feasible payoff profiles induced by behavioral strategy profiles is convex. In fact, any feasible payoff-profile can be achieved with some stage-game (behavioral) strategy profile. We then (in Lemma \ref{lem:characterization}) establish that the characterization of WRP equilibrium payoff profiles, given purely in terms of stage game strategies by \citet{maskin89}, also holds in our non-finite setting, with Lemma \ref{lem:minmax} providing the minmax payoffs needed for this characterization. We note that while Lemma 4 applies to two player games more generally, our remaining results rely, to varying degrees, on the cheap talk structure of the game.

These results allow us, for three concrete applications, to fully determine the WRP payoff profiles on the Pareto frontier by identifying the ``optimal'' strategy played by the Receiver to punish the Sender, where we can, thus, assume the Sender to be truthful.  We can, then, also fully construct the equilibrium on-path play, as well as the punishment phase play for both players.\footnote{All WRP strategies in this paper have three phases as in the useful characterization of WRP payoffs in terms of stage game strategy profiles given by \citet[Theorem 1]{maskin89}: a normal phase (on the path of play), a Sender punishment phase, (re-)triggered whenever the Sender deviates from the overall strategy, and a Receiver punishment phase, (re-)triggered whenever the Receiver deviates from the overall strategy.} Each application also has a free parameter that captures the degree of conflict, for which we can provide a comparative statics analysis.

First, we consider the perhaps simplest cheap talk stage game with binary state, message, and action spaces, akin to the judge-prosecutor example of \citet{kamenica2011bayesian}. On the path of play the Sender is truthful and the Receiver mixes between using the information fully to their own advantage and choosing the Sender-optimal action. The punishment for the Sender has the Sender still being truthful and the Receiver mixing between using the information to their own full advantage and choosing the Sender-worst action. The punishment for the Receiver has the Sender mixing between being truthful and manipulating the truth to induce the Receiver to take the Sender-optimal action and the Receiver taking the Sender's message at face value. While the maximal SPE payoff to the Receiver is independent of the probability of the conflict state, the Receiver maximal WRP payoff varies with this probability, and does so in a non-monotone fashion.  

Second, we study the non-finite quadratic preference example in \citet{crawford82} with a positive bias (the Sender always prefers higher actions than the Receiver).\footnote{The results for these games can be generalized to varying degrees.  Specifically, general distributions or somewhat more general convex Sender and Receiver preferences may be allowed.} On path play has to be such that the Receiver ultimately always implements some state-appropriate compromise action between the two players that, however, always at least slightly favors the Receiver. Of course, by our main Theorem, the Receiver can also not get any payoff close to their optimal SPE payoff. The punishments for both the Sender and Receiver involve some pooling.\footnote{Essentially, the Sender (Receiver) punishment involves pooling high (low) actions.  These resemble upper and lower censorship strategies, which are common in the literature.  We discuss this further in Section 4.2.}  While the maximal SPE payoff for the Receiver does not depend on the Sender's bias, the Receiver's maximal WRP equilibrium payoff decreases with the bias, but remains bounded away from the Receiver's minmax value (that one would get in a babbling equilibrium) as the bias tends to infinity.

Third, we consider a class of ``transparent'' games (see e.g., \citet{chakraborty2010persuasion, lipnowski2020cheap} for general transparent games) in which the Sender is risk neutral and prefers higher actions over lower actions irrespective of the state while the Receiver has quadratic preferences as in \citet{crawford82}. The parameter that we vary here is related to the variance of the state under the prior distribution. The on path play, again, has the Sender being truthful and the Receiver choosing a compromise action.

While the variance of the state under the prior does not affect the Receiver maximal SPE payoff, it does affect the Receiver maximal WRP payoff: The more spread out the prior belief about the state is (in a specific hazard rate-based stochastic dominance sense), the lower the Receiver's maximal WRP payoff (and the higher the Sender's payoff in this equilibrium). In other words, under renegotiation-proofness a higher ex-ante variance of the state hurts the Receiver and benefits the Sender. 

Let us briefly discuss the relevance of our results to the applications we mentioned above. If a politician wants to extract full information from a government official, then the politician must act in a manner that at least partially assuages the official. Similarly, if a federal government wants local governments to accurately disclose the value of a project, then the federal government may have to approve projects that are not always perfectly in line with their objectives. If a consumer wants to benefit from a repeated and trusting relationship with a seller, then they may have to pay a premium for the goods or occasionally purchase goods that are not a perfect match for the consumer's preferences. In each case there is a cost to the Receiver for their lack of information, and the Sender benefits from the information rents she captures.

Perhaps one bird's eye narrative for our results is as follows. In one-shot cheap talk games the extent of the conflict of interest between the two players clearly matters a great deal for the kind of equilibria that we can observe. In the repeated game, if we look at subgame perfection, apart from minmax payoff concerns, all of these stage game incentives play no role as there is the usual folk theorem. However, if we consider renegotiation-proofness as our notion for what constitutes credible threats, the nature and extent of the conflict of interest plays an important role in what kind of behavior we would expect to occur in situations of strategic information transmission.

While the cheap talk stage games that we allow are very general, we make strong observability (or monitoring) assumptions in the repeated game: the state is redrawn in every period, independently and identically across time, and both players can essentially observe each other's full (behavioral) strategy at the end of each stage. As strategies are mappings from states to messages and from messages to actions, respectively, it seems unlikely that such strong observability assumptions are warranted in practice. We do so, however, mostly to make our arguments simpler and to be able to utilize certain results in the literature. In Section \ref{sec:observability} we demonstrate that even when we relax these assumptions to only allow observability of the realized state, realized message, and realized action, all of our results essentially go through. The main result goes through as is, the results for the two non-finite state space applications go through as is, and other results go through for a slightly more permissive notion of renegotiation proofness.

Our main result also applies to games with large state, message, and action spaces. Many of the tools in game theory have been developed for finite games, and establishing results in games of strategic communication is often only possible when the game is finite. But many games of interest are non-finite, including the model of \citet{crawford82}, and in this sense our results are strong.

The paper is organized as follows.  We conclude the Introduction with a more detailed discussion of the related literature. Section \ref{sec:model} presents the model. Section \ref{sec:results} presents our main result. Section \ref{sec:applications} studies three specific models: a binary state model, the quadratic preference model with a continuum of states based on \citet{crawford82}, and a model with a continuum of states in which the Sender has state-independent, risk neutral preferences. Section \ref{sec:disc} discusses our results. Section \ref{sec:concl} concludes. Appendix A contains any additional or lengthy proofs.  Appendix B considers the robustness of our results to imperfect monitoring.

\subsection{Related literature}

As discussed above, there is a large literature on one-shot cheap talk games beginning with \citet{crawford82} and \citet{stokey07}.  The repeated cheap talk literature is more recent. \citet{smolin18} consider a finite cheap talk game similar to ours and establish a folk theorem. The main difference with our repeated game model is that they assume state-independent Sender preferences, monitoring is very imperfect, and uncertainty follows a Markov process. In this way they have more restrictive preferences together with a more permissive stochastic process. A paper that is closer to ours is \citet{meng21}, who essentially considers our cheap talk environment and establishes a folk theorem.  However, neither \citet{meng21} nor \citet{smolin18} consider renegotiation-proofness.\footnote{\citet{fonesca24} considers a similar problem, but with a persistent state.}

There are several other papers that consider cheap talk in dynamic environments \citep{ivanov15, ivanov16, kuvalekar22, lipnowski20,skreta14, visser17, banks00, renault13, kolotilin21, pearce19, best24}. While all of these papers consider repeated communication, they are all different from ours in various respects. Moreover, none of these papers consider the incentive problems that arise from the possibility of renegotiating previously agreed-upon equilibrium play.  

The closest of these to our environment is \citet{kuvalekar22}, though they focus on the inability to perfectly monitor behavior and do not consider renegotiation-proofness. They assume that the only public information is the history of messages, and that the history of messages alone can lead to some information transmission.

There is a literature on reputations that differs from ours by the fact that some part of the state is persistent.  \citet{sobel85} considers a model in which a Sender's preferences are uncertain, and they can be aligned or opposed. \citet{morris01} considers a simple version of such a model where a Sender does not want to be viewed as biased. \citet{ottaviani07} develop a model in which a Sender wants to be viewed as having accurate information.  In all cases the Sender cannot fully reveal their private information, and this is driven by reputation effects.  This is in contrast to our paper in which everything is driven by renegotiation incentives and full information transmission is generally possible. See \citet{strulovici22} for a discussion of renegotiation-proofness with persistent states. 

There is a large literature discussing different notions of renegotiation-proofness \citep{pearce87, maskin89, ray89, asheim91, pearce91, abreu93, krishna93, bergin93, ray94, ray97, strulovici22, pei25}.  We focus on weak renegotiation-proofness from \citet{maskin89}. Weak renegotiation-proofness is often considered a necessary condition for immunity to renegotiation, though it may be criticized as too weak. 

There are three main reasons why we focus on WRP equilibria. First, one of our main results is negative: the Receiver-optimal payoff is unachievable. Hence, this result will extend to any stronger requirement. Second, there is some disagreement as to what the appropriate strengthening of weak renegotiation-proofness is, while there is little disagreement regarding weak renegotiation-proofness as a necessary condition. Third, strengthening weak renegotiation-proofness can lead to existence issues.  Indeed, in general the set of \emph{strongly renegotiation-proof} equilibria as defined by \citet{maskin89} can be empty.\footnote{There is also some disagreement as to whether Pareto dominance is the proper requirement, but this is not an issue for two player games.}

A recent paper on advisor-advisee relationships that has a similar motivation to ours is \citet{squintani25}.  They consider the incentives of a politician to listen to one (or both) of two advisors.  One advisor is more aligned (lower bias) but less informed than the other.  A key takeaway from their results is that it is sometimes in the interest of the politician to listen to a more biased advisor.

There appears to be little work on the effect of the underlying uncertainty of the state on the equilibrium payoffs. Two clear exceptions are \citet{deimen24} who study a model with a variation of quadratic preferences, and \citet{lipnowski2020cheap} who have an example featuring convex ``V-shaped'' preferences.  In the former case uncertainty  both players' payoffs, whereas in the latter case it increases the Sender's payoffs whenever the Sender can communicate effectively. We get different results that are driven by renegotiation incentives.

\section{Model} \label{sec:model}

In this Section we present our model, along with preliminary results.  We will analyze an infinitely repeated game, and so we break the analysis up into parts.  First we describe the stage game, and then we describe the repeated game.

\subsection{The Cheap Talk Stage Game}

While we are ultimately interested in an infinitely repeated game, we begin by describing the stage game.  A general cheap talk stage game has two players, the Sender (she) and the Receiver (he). Uncertainty is described by a compact metrizable state space denoted by $\Theta$ with typical element $\theta$.  The Sender and the Receiver have a commonly known prior $\mu_0 \in \Delta(\Theta)$, where $\Delta$ applied to some set denotes the set of all probability distributions over said set.  We restrict attention to full support priors. 

Note that at this point we make no assumptions about the cardinality of $\Theta$.  If $\Theta$ is finite, then $\Delta(\Theta)$ is the set of probability distributions on $\Theta$.  If $\Theta$ is uncountable, then $\Delta(\Theta)$ is the set of probability measures associated with the Borel algebra of $\Theta$.

The Sender becomes perfectly informed after observing the state $\theta\in\Theta$. Once the Sender is informed of the state she sends a message $m$ from some set $M$, the assumed compact and metrizable message space, to the Receiver. The Receiver observes this message and chooses an action $a$ from some set $A$, the again assumed compact and metrizable action space. We will require that the message space is sufficiently rich.  More precisely, we will assume that $\Theta = M$. 

The (Bernoulli) utility functions for the Sender and Receiver, respectively, are $u_S,u_R:A \times \Theta \to \mathbb{R}$ (from action and state pairs to the real numbers) that do not depend on the message sent.  We assume that these Bernoulli utility functions are bounded and continuous in both arguments. Talk is cheap insofar as messages do not enter the payoff functions.

Stage game pure strategies are measurable functions $s:\Theta \to M$ for the Sender, and $r:M \to A$ for the Receiver. Let $\mathcal{S}$ denote the set of pure strategies for the Sender and $\mathcal{R}$ the set of pure strategies for the Receiver.   We will allow all measurable behavioral strategies. The set of behavioral strategies for the Sender and Receiver, denoted by $B(\mathcal{S})$ and $B(\mathcal{R})$, respectively, are the set of measurable functions $\sigma:\Theta\to\Delta(M)$ and $\rho:M\to \Delta(A)$, respectively. With a slight abuse of notation we denote by $\rho\circ\sigma$ both the composition of the behavioral strategies $\sigma \in B(\mathcal{S})$ and $\rho \in B(\mathcal{R})$, and the joint distribution over states and actions they induce. With another slight abuse of notation we then  denote by 
\[
u_S(\sigma,\rho)=\int_{\Theta \times A} u_S(a,\theta) d(\rho \circ \sigma)
\]
and
\[
u_R(\sigma,\rho)=\int_{\Theta \times A} u_R(a,\theta) d(\rho \circ \sigma)
\]
the (expected) utility for each behavioral strategy profile for the Sender and Receiver, respectively. Note that payoffs only depend on the composition $\rho \circ \sigma$, a fact we will sometimes utilize. 

\begin{remark} 
If we consider the set of behavioral strategies for the Sender, $B(\mathcal{S})$, modulo the equivalence $\mu_0$ almost everywhere, this set can be endowed with a topology by considering the weak topology of the joint distribution over $\Theta \times M$ that each behavioral strategy induces (all with the same $\mu_0$ marginal over $\Theta$).\footnote{See \citet{milgrom85}, who show there is a correspondence between behavioral strategies and \emph{distributional strategies}.} Analogously, the set of compositions of behavioral strategies of the form $\rho \circ \sigma$ can be endowed with the topology that is induced by the weak topology over the induced joint distributions over $\Theta \times A$. It is well-known that these sets are compact with respect to these topologies. The set of behavioral strategies for the Receiver, $B(\mathcal{R})$, does not have such a natural topology. 
\end{remark}

We will often be interested in two special (equivalence classes of) strategies.  First, we will be concerned with the \emph{truth-telling strategy}, denoted by $\tau\in \mathcal{S}$ such that $\tau(\theta)=\theta$. Second, we will be interested in the (equivalence class of) strategies of the Receiver that reflects the idea that the Receiver believes the Sender when he receives the message, for all messages.  That is, for every $m \in \Theta$ the Receiver chooses an action in the set $\mbox{argmax}_a u_R(a,m)$.\footnote{We will typically assume that this set is a singleton.} We will denote this Receiver's strategy of \emph{believing} what he is told by $\beta:M \to A$.

We end this subsection by deriving two useful Lemmas.  The first says that any payoff that can be obtained by some strategies $(\sigma,\rho)$ can also be attained by $(\tau,\rho')$ for some $\rho' \in B(\mathcal{R})$.

\begin{lemma}
\label{lem:revelation}
Every stage game payoff pair $(v_S,v_R)$ that can be achieved through some profile of behavioral strategies can also be sustained by a behavioral strategy profile in which the Sender plays $\tau$.    
\end{lemma}

\begin{proof} Let $\sigma \in B(\mathcal{S})$ and $\rho \in B(\mathcal{R}))$ be a behavioral strategy profile. Let $\tau:\Theta \to M$ be the truthful Sender strategy, i.e., $\tau(\theta)=\theta$ for all $\theta \in \Theta$. Then $(\rho, \sigma)$ and $(\tau,\rho \circ \sigma)$ induce the same joint probability distribution over states and actions.
\end{proof}

Let $\mathcal{F}=\left\{(v_S,v_R) \mid v_S=u_S(\sigma,\rho), v_R=u_R(\sigma, \rho) \mbox { for some } \sigma \in B(\mathcal{S}), \rho \in B(\mathcal{R}) \right\}$ denote the set of \emph{feasible} payoff profiles in the stage game. The following Lemma states that, in contrast to general two player games, the set $\mathcal{F}$ is convex.  

\begin{lemma} \label{lem:convexF}
For cheap talk games the set of feasible (stage game) payoff profiles $\mathcal{F}$ is convex.
\end{lemma}

\begin{proof}
This follows from Lemma \ref{lem:revelation}. To see this consider any $\sigma,\sigma' \in B(\mathcal{S})$ and any $\rho,\rho' \in B(\mathcal{R})$. We need to prove that for any $\lambda \in [0,1]$ there is a strategy profile $\sigma'',\rho''$ such that $u_i(\sigma'',\rho'')=\lambda u_i(\sigma,\rho) + (1-\lambda) u_i(\sigma',\rho')$ for both $i \in\{S,R\}$. By Lemma \ref{lem:revelation} there are $\hat{\rho},\hat{\rho}' \in B(\mathcal{R})$ such that $u_i(\sigma,\rho)=u_i(\tau,\hat{\rho})$ and $u_i(\sigma',\rho')=u_i(\tau,\hat{\rho}')$ for both $i \in \{S,R\}$. It is well known that for every behavioral strategy there is an outcome-equivalent mixed strategy. Abusing notation slightly, we let $\hat{\rho}$ and $\hat{\rho}'$ denote both the behavioral strategies as well as their outcome equivalent mixed strategies.  By the convexity of the space of mixed strategies we have that the convex combination $\lambda \hat{\rho} + (1-\lambda) \hat{\rho}'$ is also a mixed strategy and by the linearity of the (expected) payoff functions (in probabilities) we have that $\lambda u_i(\sigma,\rho) + (1-\lambda) u_i(\sigma',\rho') = u_i(\tau, \lambda \hat{\rho} + (1-\lambda) \hat{\rho}')$ for both $i \in\{S,R\}$. By the version of Kuhn's theorem given by \citet{aumann64}, there is behavioral strategy that is outcome-equivalent to mixed strategy $\lambda \hat{\rho} + (1-\lambda) \hat{\rho}'$. This concludes the proof. 
\end{proof}

\subsection{Minmax Strategies}

Next we consider minmax strategies and payoffs of the stage game.  We show that the minmax payoffs for the Receiver in cheap talk games are achieved when the Sender transmits no information, i.e., uses a constant messaging function $\sigma$. This is also the Sender's minmax strategy. For the Sender the minmax payoff is obtained when the Receiver takes his minmax strategy: he takes the action that minimizes the Sender's ex-ante payoff, regardless of the message sent. 

The minmax payoffs for the Sender and Receiver are defined as follows. 
\[
\bar{u}_S = \max_{\sigma \in B(\mathcal{S})} \min_{\rho \in B(\mathcal{R})} u_S(\sigma,\rho)
\]
and
\[
\bar{u}_R = \max_{\rho \in B(\mathcal{R})} \min_{\sigma \in B(\mathcal{S})} u_R(\sigma,\rho), 
\]
where $u_S(\sigma,\rho)$ and $u_R(\sigma,\rho)$ are the ex-ante expected payoffs induced by strategy profile $(\sigma,\rho)$.

\begin{lemma} \label{lem:minmax}
In any cheap talk game, with prior $\mu_0 \in \Delta(\Theta)$ the minmax payoffs are given by 
\[\bar{u}_R= \max_{a \in A} \mathbb{E}_{\mu_0} u_R(a,\theta)\] 
for the Receiver and 
\[\bar{u}_S= \min_{a \in A} \mathbb{E}_{\mu_0} u_S(a,\theta) \] 
for the Sender. One minmax strategy for the Sender is any constant messaging function $\sigma$. One minmax strategy for the Receiver is the constant action function $\rho(m)=a^*$ for all $m \in M$, where $a^* \in \mbox{argmin}_{a \in A} \mathbb{E}_{\mu_0} u_S(a,\theta)$.
\end{lemma}

\begin{proof}
Consider the Receiver. Trivially, the Receiver can obtain $\max_{a \in A} \mathbb{E}_{\mu_0} u_R(a,\theta)$ by playing an action in $\mbox{argmax}_{a \in A} \mathbb{E}_{\mu_0} u_R(a,\theta)$ (which exists by the compactness of $A$ and the continuity of $u_R$) regardless of the message. This implies that $\bar{u}_R \ge \max_{a \in A} \mathbb{E}_{\mu_0} u_R(a,\theta)$. To see that $\bar{u}_R$ cannot be strictly greater, suppose the Receiver uses any strategy, in which there is a message that leads to an action distribution that does not put probability one on the set of Receiver ex-ante optimal actions. Then the Sender can play this message in all states and the Receiver obtains a strictly lower payoff than if he had chosen the ex-ante optimal action. This proves the statement for the Receiver. 

Now consider the Sender. If the Sender uses a constant messaging function, then the worst that the Receiver can do to the Sender is to play an action in $\mbox{argmin}_{a \in A} \mathbb{E}_{\mu_0} u_S(a,\theta)$ (which exists by the compactness of $A$ and the continuity of $u_S$). This implies that $\bar{u}_S \ge \min_{a \in A} \mathbb{E}_{\mu_0} u_S(a,\theta)$. To see that $\bar{u}_S$ cannot be strictly greater, suppose the Sender uses any (at least partially) informative strategy, i.e., one in which different states lead to different message distributions. Then the Receiver can either still play an action in $\mbox{argmin}_{a \in A} \mathbb{E}_{\mu_0} u_S(a,\theta)$ for all messages or reduce the Sender's payoff even further by minimizing her payoff by choosing actions that minimize the Sender's payoff for every message.  
\end{proof}

\subsection{The Repeated Game}

So far we have analyzed the stage game. Now we assume that this stage game is played repeatedly at discrete points in time $t=0,1,2,...$. For simplicity, we assume that monitoring is perfect, by which we mean that after each round of play the (behavioral) strategies of both players are (essentially) fully revealed.\footnote{This assumption is extremely strong and greatly simplifies our analysis. It is not crucial for our results as we discuss in Section \ref{sec:observability}: it (essentially) suffices that players observe the realized state, the realized message, and the realized action.} Players choose strategies before the state is drawn and care about their ex-ante expected payoff in the stage game. In each round the state is redrawn from the same distribution independently of all prior draws (and all prior behavior). 

More formally, let $\pi^t$ be the pair of Sender strategy $\sigma^t$ and the composition of the period $t$ (behavioral) strategy profiles $\rho^t\circ \sigma^t$.\footnote{Defining histories like this, instead of as just paths of behavioral strategy profiles, ensures that repeated game payoffs are well-defined. This is because one can, in the spirit of \citet{milgrom85}, for each history $\pi^t$ redefine Sender strategies as the joint distributions over states and messages, and do a similar procedure for the composition of Sender and Receiver strategies as joint distributions over states and actions.} In period $t$ each player has access to the $t-1$ \emph{history} of play, given by $h^{t-1}=(\pi^0,\pi^1,...,\pi^{t-1})$.

Let $\mathcal{H}^{t}$ be the set of all possible histories up to period $t-1$, where $\mathcal{H}^0=\emptyset$. Let $\mathcal{H}=\bigcup_{t=0}^{\infty} \mathcal{H}^{t}$ denote the set of all finite histories. The repeated game strategies are, therefore, functions $\mathscr{s}:\mathcal{H} \to B (\mathcal{S})$ and $\mathscr{r}:\mathcal{H}\to B (\mathcal{R})$. Let $\mathscr{p}=(\mathscr{s},\mathscr{r})$ denote a repeated game strategy \emph{profile}.

The Sender and the Receiver evaluate a stream of (expected) stage game payoffs $u^t_S$ and $u^t_R$, respectively, for $t=0,1,2...$, by its normalized net present value, i.e. by 
\[
(1-\delta)\sum_{t=0}^{\infty} \delta^t u^t_i
\]
for $i\in\{S,R\}$. All of our upcoming results (except for the necessity part of Lemma 4, which is more general) will pertain to situations in which $\delta<1$ while being arbitrarily close to $1$.  Under these assumptions the standard folk theorem applies.  

\subsection{Weak Renegotiation-Proof Equilibrium}

A \emph{stage game (Bayes) Nash equilibrium} is a strategy profile $(\sigma^*,\rho^*)$ such that each player's strategy maximizes their payoff given the other player's strategy, i.e.,
\[
\rho^* \in \mbox{argmax}_{\rho\in B(\mathcal{R})} u_R(\sigma^*,\rho)\]
and 
\[\sigma^* \in \mbox{argmax}_{\sigma\in B(\mathcal{S})} u_S(\sigma,\rho^*).
\]

A stage game Nash equilibrium is \emph{(weak) perfect Bayesian} if in addition we assume that even for messages that are not in the support of the equilibrium Sender strategy the Receiver chooses an action that maximizes his expected utility given some (arbitrary) belief (a probability distribution) about the state. Note that if we assume that the game is such that every action $a \in A$ is optimal for the Receiver for some belief, then there is no difference between Nash and (weak) perfect Bayesian equilibria of the Sender-Receiver game.

A \emph{Nash equilibrium of the repeated game} is a repeated game strategy profile such that each player's strategy maximizes their (normalized) net present value of induced payoff streams, given the other player's strategy. A Nash equilibrium of the repeated game is a \emph{subgame perfect equilibrium (SPE)} if its induced strategy profile after each history constitutes a Nash equilibrium of the (same) repeated game. A strategy profile induced by a SPE of the repeated game at some history is called a \emph{continuation equilibrium} (of that repeated game SPE). Let the set of continuation equilibria of a repeated game SPE $\mathscr{p}=(\mathscr{s},\mathscr{r})$ be denoted by $\Sigma(\mathscr{p})$. 
 
\begin{definition}[\citet{maskin89}]
A subgame perfect equilibrium of the repeated game is \emph{weakly renegotiation-proof (WRP)} if there do not exist two continuation equilibria (of that repeated game subgame perfect equilibrium) such that one of them (strictly) Pareto-dominates the other.  
\end{definition}

Every repeated game (for which the stage game has a Nash equilibrium) has a weakly renegotiation-proof equilibrium - playing a stage game Nash equilibrium after all histories is trivially WRP. The following Lemma is a version of a characterization for two player games of WRP equilibria only in terms of stage game payoffs given by \citet{maskin89}, adapted to the class of cheap talk games.

\begin{lemma} \label{lem:characterization}
Consider a payoff profile $v=(v_S,v_R)$. There is a $\bar{\delta}<1$ such that for all $\delta \in (\bar{\delta},1)$ payoff profile $v$ is WRP if $v$ is feasible and strictly individually rational, $v_S > \bar{u}_S, v_R > \bar{u}_R$, and there are (behavioral) strategy profiles $(\sigma^S,\rho^S)$ and $(\sigma^R,\rho^R)$ (in the stage game) such that
\[
u_R(\sigma^S,\rho^S) \ge v_R \mbox{ and } \sup_{\sigma' \in B (\mathcal{S})} u_S(\sigma',\rho^S) < v_S 
\]
and
\[
u_S(\sigma^R,\rho^R) \ge v_S \mbox{ and } \sup_{\rho' \in B (\mathcal{R})} u_R(\sigma^R,\rho') < v_R.
\]
For $\delta\in(0,1)$ a necessary condition for payoff profile $v=(v_S,v_R)$ to be WRP is that all of the above inequalities hold weakly.
\end{lemma}

The proof is in the Appendix.  It helps for building intuition to think of $(\sigma^S,\rho^S)$ as the strategy profile in the stage game that is intended as the punishment for the Sender, and $(\sigma^R,\rho^R)$ as the punishment for the Receiver. Note that in any punishment the punisher needs to achieve a payoff at least as high as on the equilibrium path of play. 

\begin{remark}[On the sufficiency part of Lemma \ref{lem:characterization}] \label{rem:necessitypartFM}
\citet{maskin89} provide this as their Theorem 1 for the mixed extension of arbitrary finite two player games (with observable mixed actions). For that class of games \citet{gunther2019note} have shown that the proof for the sufficiency part provided by \citet{maskin89} has one incorrect (or missing) step. \citet{gunther2019note} provide a proof of a slightly weaker version of that step, with the result that for the mixed extension of arbitrary finite two player games the sufficient condition for \citet{maskin89} Theorem is that at least one of the two weak inequalities holds strictly. The difficulty arises if there are repeated game payoff profiles that cannot be sustained by a mixed strategy profile (with the two players mixing independently).

For the present case of the class of arbitrary cheap talk games (with observable behavioral stage game strategies) the original \citet{maskin89} proof holds. This is so because the set of payoff profiles sustainable by independent mixtures over Sender and Receiver pure strategies is a convex set by Lemma \ref{lem:convexF}. Therefore, all feasible repeated game payoff profiles can be sustained with a stage game behavioral strategy profile (the same profile played at all histories). 
\end{remark}

\begin{remark}[On the necessity part of Lemma \ref{lem:characterization}]\label{rem:necessity2partFM}\citet{maskin89} prove their Theorem 1 for finite games. Finite games have compact mixed strategy spaces and this is an aspect that \citet{maskin89} use to prove their necessary condition.  In Appendix \ref{app:necessity} we provide a more direct proof of the necessary condition without relying on the compactness assumption that is of some independent interest as the proof of the necessity part of Lemma \ref{lem:characterization} does not require the special structure of cheap talk games and applies to all two player games (with or without compact strategy spaces). Note also, as in \citet{maskin89}, that the necessary conditions are necessary for all discount factors $\delta \in(0,1)$, not just for the limit as $\delta \to 1$. 
\end{remark}

To study WRP in cheap talk games, one can restrict attention to Sender punishments $(\sigma^S,\rho^S)$, where $\sigma^S=\tau$, i.e. the Sender is truthful.  This is established in the following Lemma.

\begin{lemma} 
\label{lem:worstpunishmentsender}
Let $(\sigma,\rho)$ be any behavioral stage-game strategy profile. There is a $\rho' \in B(\mathcal{R})$ such that 
\[
u_R(\tau,\rho') = u_R(\sigma,\rho) \mbox{ and } \sup_{\sigma' \in B(\mathcal{S})} u_S(\sigma',\rho') \le \sup_{\sigma' \in B(\mathcal{S})} u_S(\sigma',\rho). 
\]
\end{lemma}

Lemma \ref{lem:worstpunishmentsender} is used in Section \ref{sec:applications} to determine the upper bound of the Receiver payoff in any WRP on the Pareto frontier in two applications.

\begin{proof}
Let $\sigma \in B(\mathcal{S})$ and $\rho \in B(\mathcal{R})$ be a behavioral stage-game strategy profile satisfying the conditions of the Lemma. By Lemma \ref{lem:revelation}, there is a $\rho' = \rho \circ \sigma \in B(\mathcal{R})$ such that $u_R(\tau,\rho') = u_R(\sigma,\rho)$. Then for any $\sigma'' \in B(\mathcal{S})$, $\sigma \circ \sigma'' \in B(\mathcal{S})$ and we have $u_S(\sigma'',\rho \circ \sigma)=u_S(\sigma \circ \sigma'', \rho)$. In words, any payoff the Sender can achieve against $\rho \circ \sigma$ the Sender can also achieve against $\rho$. Therefore $\sup_{\sigma' \in B(\mathcal{S})} u_S(\sigma',\rho') \le \sup_{\sigma' \in B(\mathcal{S})} u_S(\sigma',\rho)$.

\end{proof}

\section{Main Result} \label{sec:results}

Now we present our main result, which requires two additional assumptions. To state these, let $A^* =\left\{a \in A \mid a \in \mbox{argmax}_{a'} u_R(a',\theta) \mbox{ for some } \theta \in \Theta \right\}$ be the set of actions that are optimal for the Receiver for some state.  That this set is well-defined follows from the compactness of $A$ and the continuity of $u_R$. Furthermore, let $a^*(\theta)$ denote, for all $\theta \in \Theta$, the set of Receiver-optimal actions in state $\theta$. 

The first assumption states that the Receiver has a unique and distinct optimal best response action in each state. 

\begin{assumption}[Receiver unique best response] \label{ass:uniquebest}
For every $\theta \in \Theta$ the set of Receiver best response actions $a^*(\theta)$ is a singleton. Moreover if $a \in a^*(\theta)$ for some $\theta \in \Theta$, then $a \not\in a^*(\theta')$ for any $\theta' \in \Theta$ such that $\theta' \ne \theta$.
\end{assumption}

Under this assumption $a^*(\theta)$ is a one-to-one function from $\Theta$ to $A^*$. The second assumption requires that there is at least some degree of conflict of interest between the Sender and the Receiver.

\begin{assumption}[Minimal Conflict] \label{ass:minconf}
There exists a (measurable) $Z \subseteq \Theta$ with $\mu_0(Z)>0$ such that $u_S(a^*(\theta),\theta) < \max_{a\in A^*} u_S(a,\theta)$ for all $\theta\in Z$.\footnote{Note that  the Sender maximizes only over the set $A^*$ in this definition. This is necessary, because the Sender can typically only induce actions in $A^*$.  We also note that by continuity of $u_R$ and compactness of $A$, the set $A^*$ is compact, and so this maximum exists.}
\end{assumption}

Recall that the standard folk theorem states that any feasible and individually rational (higher than minmax payoffs) payoff profile can be sustained by a subgame perfect equilibrium of the repeated game for sufficiently high discount factors. We can now state the main result.

\begin{theorem} \label{thm:main}
Consider any cheap talk game with full support prior (and continuous Bernoulli utility functions for both players).  Under Assumptions \ref{ass:uniquebest} (Receiver unique best responses) and \ref{ass:minconf} (minimal conflict) there exists an $\eta >0$ such that for all $\delta \in (0,1)$ all WRP equilibrium payoff profiles of the $\delta$-discounted infinitely repeated game $(v_S,v_R)$ satisfy $v_R\le u_R(\tau,\beta)-\eta$. 
\end{theorem}

In other words, Theorem \ref{thm:main} states that there is no WRP equilibrium that provides the Receiver arbitrarily close to his optimal SPE payoff.  The proof is in the Appendix, but we provide a brief sketch here.

First we show that the Receiver-optimal SPE is not WRP.  The Receiver-optimal SPE can only be achieved by essentially one strategy pair: $(\tau,\beta)$ (or outcome equivalent transformations thereof).  But (in all these cases), given the Assumption of minimal conflict, the Sender can increase her payoff by lying, which will violate the condition in Lemma \ref{lem:characterization}. 

Next we show in Lemma \ref{lem:cocont} that if a strategy profile gives the Receiver a payoff close to the Receiver-optimal payoff, then two things must be true.  First, the Sender's payoff at this profile must be close to her payoff at the the Receiver-optimal strategy.  Second, the Sender's best response to this profile cannot be much worse than her best response to the Receiver-optimal strategy.  This allows us to derive a contradiction via Lemma \ref{lem:characterization} ensuring the payoff cannot be WRP.

\begin{remark}
If there is no conflict of interest between the Sender and the Receiver, then $(\tau,\beta)$ is a Nash equilibrium of the stage game and always playing this after any history is trivially subgame perfect as well as WRP, as there are simply no incentives to deviate for either party.
\end{remark}

The following Example shows that Theorem \ref{thm:main} may fail to hold if the Receiver unique best response Assumption \ref{ass:uniquebest} is violated. 

\begin{example} \label{ex:counter}

Let $\Theta=\{0,1\}=M$ and $A=\{0,1,2,3\}$, with common prior $\mu_0=\frac{1}{3}$ (the probability of the state being $1$). The Sender and Receiver payoffs are given in Table \ref{tab:payoffscounter}. Note that the Receiver is indifferent between actions 0 and 3, and actions 1 and 2. However, the Sender sometimes receives a large negative payoff from actions 2 and 3.  In this way, the Receiver essentially has access to punishments that are costless to the Receiver but costly to the Sender.

To see that the Receiver-optimal can be implemented we assign punishment strategies. The on path play has the Sender truthful (playing $\tau$) and the Receiver playing 0 if 0 is reported and 1 if 1 is reported. If the Sender deviates, the continuation strategy has the Sender playing $\tau$ and the Receiver playing 3 if 0 is reported and 2 if 1 is reported.  In each case, the Sender's best response payoff is $0<\frac{1}{3}$ and hence the profile is WRP.

\end{example}

\begin{table}[htbp] 
\begin{center}
\begin{tabular}{c|ccccc}
state $\backslash$ action & \multicolumn{1}{c}{0} & \multicolumn{1}{c}{1}& \multicolumn{1}{c}{2}& \multicolumn{1}{c}{3}\\
\hline
0 & 0,1 & 1,0 & 0,0 &-10,1 \\
1 & 0,0 & 1,1 & -10,1 & 0,0 \\
\end{tabular}
\caption{Payoffs (first Sender, second Receiver) in the stage game from Example \ref{ex:counter}.}
\label{tab:payoffscounter} 
\end{center} 
\end{table}

Theorem \ref{thm:main} provides a negative result, in that it states what cannot be a WRP equilibrium. This would not be very informative if there were no WRP equilibrium at all. Playing the stage game (perfect Bayesian) Nash equilibrium after every history is always WRP, but stage game equilibria will typically not be on the Pareto frontier. A natural question is whether a WRP equilibrium exists on the Pareto frontier.  Unfortunately, we cannot establish this affirmatively in general. However, \citet{evans89} have shown that for any generic finite stage game the repeated game has a WRP equilibrium on the Pareto frontier. More precisely their result implies that for any finite cheap talk game there exists a ``nearby'' game (i.e. a game obtained by perturbing the payoffs slightly) that contains a WRP Pareto efficient equilibrium.  One might be concerned that such a perturbed game need not be a cheap talk game (i.e., $m$ may directly enter the utility functions). However, an examination of their proof indicates that this is not the case; one can qualify genericity as being within the class of cheap talk games.

In the next section we use Lemma \ref{lem:worstpunishmentsender} in three applications of some independent interest to provide a sharp analytical characterization of Receiver maximal WRP payoffs. In all three applications, when we identify WRP payoff profiles for $\delta \to 1$ we mean that for each WRP payoff profile there is a $\bar{\delta}<1$ such that for all $\delta \in (\bar{\delta},1)$ there is a WRP equilibrium of the repeated game that implement this payoff profile. The first application is the simplest binary game of interest (akin to the prosecutor judge example of \citet{kamenica2011bayesian}). The second is the prime example of the classical paper on cheap talk, \citet{crawford82}. The third is a class of cheap talk games with state-independent, risk neutral Sender preferences. 

\section{Applications} \label{sec:applications}

\subsection{A Binary Game} \label{sec:binary}

In this section we study the perhaps simplest problem of interest, in which the politician (more generally the Receiver) asks the government official (more generally the Sender) simple true/false or yes/no questions. The Receiver would like to know the truth and adapt his chosen action accordingly and the Sender would like the Receiver to always choose the same action irrespective of the true state. 

More formally, the Sender observes a state $\theta\in\Theta=\{0,1\}$, and communicates to the Receiver through a message $m\in M=\{0,1\}$. The Sender chooses a message-function $\sigma:\Theta\to \Delta(M)$, or in this case $\sigma:\{0,1\}\to \Delta (\{0,1\})$.  The Receiver chooses an action $a\in A=\{0,1\}$. The Receiver chooses a strategy denoted $\rho:M\to \Delta (A)$, or in this case $\rho:\{0,1\}\to\Delta(\{0,1\})$, and receives ex post utility $1-(a-\theta)^2$. The Sender receives ex post utility $a$. The stage game payoffs are also given in Table \ref{tab:payoffssimple}. The two agents have a common prior probability $\alpha \in [0,1]$ that $\theta=0$.

\begin{table}[htbp]
\begin{center}
\begin{tabular}{c|ccc}
state $\backslash$ action & \multicolumn{1}{c}{0} & \multicolumn{1}{c}{1}  \\
\hline
0 & 0,1 & 1,0  \\
1 & 0,0 & 1,1  \\
\end{tabular}
\caption{Payoffs in the stage game as a function of the state (row) and the action taken by the Receiver (column). The first entry is the Sender's payoff, the second entry is the Receiver's payoff.}
\label{tab:payoffssimple} 
\end{center} 
\end{table}

\begin{table}[htbp]
\centering
\begin{tabular}{c|cccc}
S $\backslash$ R & always 0 & believe & opposite & always 1 \\
\hline
always 0 & \textcolor{red}{$0,\alpha$} & $0,\textcolor{red}{\alpha}$ & $\textcolor{red}{1},1-\alpha$ & $\textcolor{red}{1},1-\alpha$ \\
truthful & $\textcolor{red}{0},\alpha$ & $1-\alpha,\textcolor{red}{1}$ & $\alpha,0$ & $\textcolor{red}{1},1-\alpha$ \\
opposite & $\textcolor{red}{0},\alpha$ & $\alpha,0$ & $1-\alpha,\textcolor{red}{1}$ & $\textcolor{red}{1},1-\alpha$ \\
always 1 & \textcolor{red}{$0,\alpha$} & $\textcolor{red}{1},1-\alpha$ & $0,\textcolor{red}{\alpha}$ & $\textcolor{red}{1},1-\alpha$
\end{tabular}       
\caption{The stage game in ex-ante normal form. Best responses, here assuming $\alpha \ge \frac12$, are in \textcolor{red}{red}.}
\label{tab:payoffssimpleNF} 
\end{table}    

Each player has four pure strategies, as given in the ex-ante normal form matrix representation of this game in Table \ref{tab:payoffssimpleNF}. Recall that we allow players to randomize. 

\begin{figure}[htb]
\centering
\begin{tikzpicture}[scale=1.6]

\draw [<->] (0,6.5) node (yaxis) [above] {$u_R$}
        |- (6.5,0) node (xaxis) [right] {$u_S$};
\draw (0,6) node[left] {$1$};
\draw (6,0) node[below] {$1$};
\draw (0,0) node[below] {$0$} node[left] {$0$};

\draw (2,0) node[below] {$\frac{1}{3}$};
\draw (4,0) node[below] {$\frac{2}{3}$};

\draw (0,2) node[left] {$\frac{1}{3}$};
\draw (0,4) node[left] {$\frac{2}{3}$};
\draw (0,4) -- (2,6) -- (6,2) -- (4,0) -- (0,4);

\filldraw [fill=white,draw=red] (4,4) circle (2pt);

\filldraw [fill=white,draw=red] (3,5) circle (2pt);

\draw[dotted] (0,4) -- (6,4);

\filldraw [black] (0,4) circle (2pt);
\filldraw [black] (2,6) circle (2pt);
\filldraw [black] (6,2) circle (2pt);
\filldraw [black] (4,0) circle (2pt);

\draw [red, very thick] (3.05,4.95) -- (3.95,4.05);
\draw (3.5,4.7) node[right] {WRP};
\end{tikzpicture}
\caption{The set of payoff profiles for the binary for the case of $\alpha=\frac23$. The dotted line is the Receiver's minmax payoff of $\frac23$. The Sender's minmax payoff is $0$.  The red line segment illustrates the Pareto efficient WRP equilibria for $\delta \to 1$.}
\label{fig:payoffsimple} 
\end{figure}

Assume that $\alpha \ge \frac12$ (the case of $\alpha < \frac12$ is similar). Note that the only (Bayes) Nash equilibria of this stage game are babbling equilibria, such as the Sender always sending message $1$ and the Receiver always playing action $0$ with payoffs of $0$ for the Sender and $\alpha$ for the Receiver. Note, furthermore, that this game has a social dilemma aspect in that the unique stage game Nash equilibrium payoff profile of $(0,\alpha)$ is Pareto-dominated by the payoff profile induced by the Sender being truthful and the Receiver believing her. 

\begin{figure}[htb]
\centering

% First subfigure
\begin{subfigure}{0.45\textwidth}
\centering
\begin{tikzpicture}[scale=0.7]

\draw [<->] (0,6.5) node (yaxis) [above] {$u_R$} |- (6.5,0) node (xaxis) [right] {$u_S$};
\draw (0,6) node[left] {$1$};
\draw (6,0) node[below] {$1$};
\draw (0,0) node[below] {$0$} node[left] {$0$};

\draw (1,0) node[below] {$\frac{1}{6}$};
\draw (5,0) node[below] {$\frac{5}{6}$};

\draw (0,1) node[left] {$\frac{1}{6}$};
\draw (0,5) node[left] {$\frac{5}{6}$};
\draw (0,5) -- (1,6) -- (6,1) -- (5,0) -- (0,5);

\filldraw [fill=white,draw=red] (2,5) circle (2pt);

\filldraw [fill=white,draw=red] (1.715,5.285) circle (2pt);

\draw[dotted] (0,5) -- (6,5);

\filldraw [black] (0,5) circle (2pt);
\filldraw [black] (1,6) circle (2pt);
\filldraw [black] (6,1) circle (2pt);
\filldraw [black] (5,0) circle (2pt);

\draw [red, very thick] (1.765,5.235) -- (1.95,5.05);
\draw (1.8,5.25) node[right] {WRP};
\end{tikzpicture}
\caption{$\alpha=\frac56$.}  
\label{subfig:56}
\end{subfigure}
\hfill
% Second subfigure
\begin{subfigure}{0.45\textwidth}
\centering
\begin{tikzpicture}[scale=0.7]

\draw [<->] (0,6.5) node (yaxis) [above] {$u_R$}
        |- (6.5,0) node (xaxis) [right] {$u_S$};
\draw (0,6) node[left] {$1$};
\draw (6,0) node[below] {$1$};
\draw (0,0) node[below] {$0$} node[left] {$0$};

\draw (2.5,0) node[below] {$\frac{5}{12}$};
\draw (3.5,0) node[below] {$\frac{7}{12}$};

\draw (0,2.5) node[left] {$\frac{5}{12}$};
\draw (0,3.5) node[left] {$\frac{7}{12}$};
\draw (0,3.5) -- (2.5,6) -- (6,2.5) -- (3.5,0) -- (0,3.5);

\filldraw [fill=white,draw=red] (5,3.5) circle (2pt);

\filldraw [fill=white,draw=red] (3.5295,4.9705) circle (2pt);

\draw[dotted] (0,3.5) -- (6,3.5);

\filldraw [black] (0,3.5) circle (2pt);
\filldraw [black] (2.5,6) circle (2pt);
\filldraw [black] (6,2.5) circle (2pt);
\filldraw [black] (3.5,0) circle (2pt);

\draw [red, very thick] (3.5795,4.9205) -- (4.95,3.55);
\draw (4.35,4.55) node[right] {WRP};
\end{tikzpicture}
\caption{$\alpha=\frac{7}{12}$.} 
\label{subfig:712}
\end{subfigure}

\vspace{1em}

% Third subfigure
\begin{subfigure}{0.45\textwidth}
\centering
\begin{tikzpicture}[scale=0.7]

\draw [<->] (0,6.5) node (yaxis) [above] {$u_R$}
        |- (6.5,0) node (xaxis) [right] {$u_S$};
\draw (0,6) node[left] {$1$};
\draw (6,0) node[below] {$1$};
\draw (0,0) node[below] {$0$} node[left] {$0$};

\draw (3,0) node[below] {$\frac{1}{2}$};

\draw (0,3) node[left] {$\frac{1}{2}$};

\draw (0,3) -- (3,6) -- (6,3) -- (3,0) -- (0,3);

\filldraw [fill=white,draw=red] (6,3) circle (2pt);

\filldraw [fill=white,draw=red] (4,5) circle (2pt);

\draw[dotted] (0,3) -- (6,3);

\filldraw [black] (0,3) circle (2pt);
\filldraw [black] (3,6) circle (2pt);

\filldraw [black] (3,0) circle (2pt);

\draw [red, very thick] (4.05,4.95) -- (5.95,3.05);
\draw (5.55,4.05) node[right] {WRP};
\end{tikzpicture}
\caption{$\alpha=\frac{1}{2}$.} 
\label{subfig:12}
\end{subfigure}
\hfill
% Fourth subfigure
\begin{subfigure}{0.45\textwidth}
\centering

\begin{tikzpicture}[scale=0.7]

\draw [<->] (0,6.5) node (yaxis) [above] {$u_R$}
        |- (6.5,0) node (xaxis) [right] {$u_S$};
\draw (0,6) node[left] {$1$};
\draw (6,0) node[below] {$1$};
\draw (0,0) node[below] {$0$} node[left] {$0$};

\draw (2,0) node[below] {$\frac{1}{3}$};
\draw (4,0) node[below] {$\frac{2}{3}$};

\draw (0,2) node[left] {$\frac{1}{3}$};
\draw (0,4) node[left] {$\frac{2}{3}$};
\draw (0,2) -- (4,6) -- (6,4) -- (2,0) -- (0,2);

\filldraw [fill=white,draw=red] (6,4) circle (2pt);

\filldraw [fill=white,draw=red] (4.8,5.2) circle (2pt);

\draw[dotted] (0,4) -- (6,4);

\filldraw [black] (0,2) circle (2pt);
\filldraw [black] (4,6) circle (2pt);

\filldraw [black] (2,0) circle (2pt);

\draw [red, very thick] (4.85,5.15) -- (5.95,4.05);
\draw (5.55,4.55) node[right] {WRP};
\end{tikzpicture}
\caption{$\alpha=\frac{1}{3}$.}  
\label{subfig:13}
\end{subfigure}
\caption{The set of payoff profiles for the binary game for different degrees of conflict, $\alpha$. The set of WRP payoff profiles for $\delta \to 1$ on the Pareto frontier is given in red.}
\label{fig:binaryother}
\end{figure}

For the game at hand we have minmax payoffs of $\bar{u}_R=\max\{\alpha,1-\alpha\}$ and $\bar{u}_S=0$. The standard folk theorem then states that for every payoff profile $(v_S,v_R)$ in the convex hull of stage game payoff profiles induced by pure strategy profiles of the stage game (see Figure \ref{fig:payoffsimple}) that exceeds the minmax payoff profile $(\bar{u}_S,\bar{u}_R)$, there is a discount factor $\bar{\delta} < 1$ such that for all $\delta \ge \bar{\delta}$ there is a subgame perfect equilibrium of the repeated game that induces payoff profile $(v_S,v_R)$.

In particular, the Receiver-maximal payoff of $1$ can be sustained in a subgame perfect equilibrium of the repeated game. This is the equilibrium that the proverb, ``Fool me once, shame on you. Fool me twice, shame on me," seems to refer to.  

Now let us examine the WRP equilibria (for $\delta \to 1$) of this repeated game.  To see that the payoff profile $(1-\alpha,1)$ (i.e. the Receiver-optimal payoff profile) is not WRP, we can use Lemma \ref{lem:characterization}. Suppose otherwise, that it is WRP. Then $(\sigma^S,\rho^S)$ must satisfy $u_R(\sigma^S,\rho^S) \ge 1$. There is only one strategy profile (two if we allow opposite strategies) that provides such a payoff. Therefore, $(\sigma^S,\rho^S)$ must be the strategy profile in which the Sender tells the truth and the Receiver believes her. But then $\max_{\sigma \in B(\mathcal{S})} u_S(\sigma,\rho^S)=1 > 1-\alpha$, a contradiction. 

Any payoff profile on the Pareto frontier is WRP if and only if it satisfies that the Receiver payoff is in the interval $\left(\max\{\alpha,1-\alpha\},1-\frac{\alpha(1-\alpha)}{2-\alpha}\right)$. The upper bound is less than $1$ as long as $\alpha < 1$. For the case of $\alpha=\frac23$, for instance, the interval becomes $(\frac23,\frac56)$ as indicated in Figure \ref{fig:payoffsimple}. We formalize this idea in the following Proposition.

\begin{proposition} \label{prop:binary}
Consider the binary game with $\alpha \in (0,1)$. A payoff profile on the Pareto frontier is WRP for $\delta \to 1$ if and only if the Receiver's payoff is in $\left(\max\{\alpha,1-\alpha\},1-\frac{\alpha(1-\alpha)}{2-\alpha}\right)$. These payoffs can be sustained by the following strategy profile. On the path of play the Sender is truthful, and the Receiver mixes appropriately between believing and always 1 (the most preferred action of the Sender). To obtain the Receiver maximal WRP payoff for $\delta \to 1$ the Receiver plays believe with probability $\frac{1}{2-\alpha}$ and always 1 with probability $\frac{1-\alpha}{2-\alpha}$. The punishment for the Sender has the Sender be truthful, and the Receiver mixing appropriately between believing and always playing 0 (the least preferred action of the Sender). The punishment for the Receiver has the Sender mixing between being truthful and always sending 1 and the Receiver believing.
\end{proposition}

The proof is in the Appendix. Note that $\alpha$ is the probability of state $0$, that is, the probability of the conflict state. In this sense, $\alpha$ is a measure of the degree of conflict of interest between the two players. Note that the maximal Receiver payoff in any WRP (on the Pareto frontier) varies non-monotonically with the degree of conflict $\alpha$ as also illustrated in Figure \ref{fig:binarycompstat}.

\begin{figure}[htb]
\centering
	\begin{tikzpicture}[scale=0.9]
	\begin{axis}[
		axis lines=middle,
		xlabel={$\alpha$},
		ylabel={$u_R$},
		xmin=0, xmax=1,
		ymin=0.5, ymax=1.05,
		samples=200,
		domain=0:1,
		width=14cm,
		height=10cm
		]
		
		\addplot[
		blue,
		thick
		]
		{
			max(x,1-x)
		};
		
		\addplot[
		blue,
		thick
		]
		{
			1-(x*(1-x))/(2-x)
		};
		
		\addplot[
		red,
		thick
		]
		{
			1/2 + (1/2)*max(x,1-x)
		};
		
	\end{axis}
\end{tikzpicture}	
    \caption{The payoff of the Receiver in the binary game, as a function of the degree of conflict $\alpha$, in a WRP (when $\delta \to 1$) on the Pareto frontier can be anywhere between the two blue lines. The lower bound is the Receiver's minmax value $\max(\alpha,1-\alpha)$, the upper bound is the one given in Proposition \ref{prop:binary}. The red line indicates the halfway point between the Receiver's maximal SPE payoff of $1$ and his minmax payoff. Note that for some $\alpha$ the upper bound is below and for some above the red line.}
    \label{fig:binarycompstat}
\end{figure}

\subsection{Quadratic Preferences} \label{sec:cs}

In this Section we study the well-known environment from \citet{crawford82}. We will focus on the special case of the quadratic utilities and uniform distribution, but some of our results generalize a bit. The politician (Receiver) must choose a policy $a\in A=[-b,1+b]$ where $b$ is non-negative. The Receiver wants to choose a state-appropriate action for each state. Somewhat simplified, the Receiver wants to match his action to the state $\theta\in\Theta=[0,1]$, more specifically his ex post utility is $-(a-\theta)^2$. 

The government official (Sender) becomes informed of the state $\theta$. Ex-ante, the two players have a common (full support) prior denoted by $\mu_0$, assumed to be the uniform distribution over $[0,1]$. There is a conflict of interest: the Sender has an ex post utility of $-(a-b-\theta)^2$. That is, given the state, the official's ideal policy is higher (by $b$) than the politician's ideal policy.

For this model, \citet{crawford82} show that full information transmission can occur in equilibrium (of the stage game) if and only if $b=0$.  However, unlike in the binary example, some information can be transmitted in equilibrium if $b>0$ is sufficiently small.\footnote{In this case there can be a multiplicity of equilibria.}

\begin{figure}[htb]
\centering
\begin{tikzpicture}[scale=1.8]

\draw [->] (6,1) -- (6,6.5) node [above] {$u_R$};
\draw [->] (0,6) -- (6.5,6) node [right] {$u_S$};

\draw (6,3) arc (0:90:3);
\filldraw [black] (6,3) circle (2pt);
\filldraw [black] (3,6) circle (2pt);
\draw (6,3) node [right] {$-b^2$};
\draw (3,6) node [above] {$-b^2$};

\filldraw [fill=white,draw=red, thick] (5.12,5.12) circle (2pt);

\filldraw [fill=white,draw=red, thick] (3.98,5.83) circle (2pt);

\draw [red, ultra thick] (5.075,5.175) arc (46.5:70:3);

\draw (4.6,5.2) node[left] {WRP};

\draw[dotted] (0,2) -- (6,2) node [right] {$-V$};

\draw[dotted] (1,1) -- (1,6) node [above] {$-V-B^2$};
\end{tikzpicture}
\caption{Pareto frontier (and WRP payoffs for $\delta \to 1$ on that frontier) of the \citet{crawford82} Example.}
\label{fig:cs}
\end{figure}

For this game, the minmax payoffs are $\bar{u}_R=-V$ (action $a=\mathbb{E}_{\mu_0}[\theta]$) and $\bar{u}_S=-V-\left(\mathbb{E}_{\mu_0}[\theta] + 2b \right)^2$ (action $a=-b)$, where $V$ is the variance of $\theta$, i.e. $V=\mathbb{E}_{\mu_0}[\theta-\mathbb{E}_{\mu_0}[\theta]]^2$. One could call $B:=\mathbb{E}_{\mu_0}[\theta] + 2b$ the expected ``bias'' (as in statistics -- here the difference between desired action 1 and implemented action $-b$) for the Sender.  

The Pareto frontier of feasible payoff profiles is sustained by the set of strategy profiles that solve
\[
\max_{\sigma\in B(\mathcal{S}),\rho\in B(\mathcal{R})}\lambda u_S(\sigma,\rho)+(1-\lambda)u_R(\sigma,\rho)
\]
for $\lambda\in[0,1]$. For the game at hand we can trace out the Pareto frontier as all those points $(v_S,v_R)$ that satisfy $\sqrt{-v_S}+\sqrt{-v_R}=b$, which we have sketched in Figure \ref{fig:cs}.\footnote{The Pareto frontier in this case is the bottom left quarter of the $L^{\frac12}$ circle (i.e., the circle defined using the $L^{\frac12}$-quasinorm).} To implement a point on the Pareto frontier, corresponding to some $\lambda \in [0,1]$, the Sender has to be truthful (or play any strictly monotone strategy) and the Receiver has to choose action $\theta + \lambda b$ for all $\theta \in \Theta$. For $\lambda=0$ the Receiver's chosen action is the Receiver optimal action, for $\lambda=1$ it is the Sender optimal action and for $\lambda \in (0,1)$ it is a compromise action. The lower $\lambda$ the more the compromise action favors the Receiver. 

In Figure \ref{fig:cs} we have sketched the Pareto frontier and the subset of WRP payoff profiles that are on the Pareto frontier.

\begin{proposition} \label{prop:CS}
Consider the model with quadratic preferences. Any WRP payoff profile $(v_S,v_R)$ for $\delta \to 1$ on the Pareto frontier satisfies $v_S\le v_R$. On the equilibrium path, the Receiver chooses action $\theta + \lambda b$ for all $\theta \in \Theta$ for some $\lambda \le \frac12$, i.e., the Receiver chooses a compromise action that favors the Receiver.  
\end{proposition}

The proof is in the Appendix.  One way to read Proposition \ref{prop:CS} is this: one could think of the Pareto frontier as the range of possible outcomes of some form of bargaining between the Sender and the Receiver. Then Proposition \ref{prop:CS} states that any WRP profile (the outcome of this bargaining through repeated cheap talk) always favors the Receiver in that the Receiver receives more of the surplus that truth-telling induces than the Sender does. Of course, by Theorem \ref{thm:main} the Receiver cannot obtain all of the surplus.

In the Appendix we, actually, prove a more general version in which the prior has full support but is otherwise arbitrary and $u_R(a,\theta)=f(a-\theta)$ and $u_S(a,\theta)=f(a-\theta-b)$, where $f$ is continuous, twice differentiable, strictly concave, and takes a maximum at $f(0)=0$, and the prior has full support. Hence, if the Sender and Receiver have the same maximum utility (here normalized to 0) and the same ``loss function'' $f$, then the Sender can never get more than the Receiver. Also in this case, this means that, on the equilibrium path, the Receiver plays a compromise action that favors the Receiver.

However, this result does not fully generalize further.  For an example, notice that the binary game of Section 4.1 can be rewritten as a special case of the discretized continuum game with a state-dependent Sender bias: The Receiver has the same utility function as in the continuum game; the Sender's utility can be written as $u_S(a,\theta)=1-(a-\theta-b(\theta))^2$, with state dependent bias $b(0)=1$ and $b(1)=0$. In this case, the Sender can get a higher payoff than the Receiver, for instance, if $\alpha<\frac{2}{3}$.

The following proposition provides some facts about the set of WRP equilibria with payoffs on the Pareto frontier for the continuum game for any bias $b>0$ (and by symmetry for any $b <0$).\footnote{For $b=0$ the game also has a trivial WRP: there is no conflict of interest and the Receiver- and at the same time Sender-optimal payoff profile is WRP as there are simply no incentives to deviate from it.} 

\begin{proposition} \label{prop:CSexist}
Consider the model with quadratic preferences and a uniform distribution. For any bias $b>0$ there is a $\bar{\lambda}_b \in(0,\frac12)$ such that a payoff profile $(v_S,v_R)$ is WRP for $\delta \to 1$ and on the Pareto frontier if and only if $v_S=-(1-\lambda)^2 b^2$ and $v_R=-\lambda^2 b^2$ for some $\lambda \in (\bar{\lambda}_b,\frac12)$. The Receiver's supremum WRP payoff for $\delta \to 1$ is strictly decreasing in $b$ and converges to $-\frac{(1-2^{-2/3})^3}{3} \approx -0.01689$ as $b \to \infty$ (which is substantially higher than $-1/12$, the Receiver's minmax value equal to the variance of the uniform distribution).  
\end{proposition}

Some aspects of Proposition \ref{prop:CSexist} is given in Figures \ref{fig:WRPmaxCS} and \ref{fig:WRPmaxCSlambda}. The proof is constructive and given in the Appendix. Here we provide a brief sketch of the proof. For the Receiver punishment the Sender pools the states on some interval $[0,y]$ and is fully revealing otherwise, while the Receiver plays the Sender-optimal (given his information). Note that varying $y$ traces the payoffs from pure babbling to fully revealing, the Receiver can always obtain the Sender's payoff, and payoffs are continuous in $y$.  As a consequence, one can find a $y$ that punishes the Receiver to support a WRP equilibrium for all $\lambda\in(0,\frac{1}{2})$. The Receiver punishment does, therefore, not limit the range of WRP payoffs beyond what is stated in Proposition \ref{prop:CS}.

We then identify among all punishments for the Sender that keep the Sender supremum payoff fixed, the one that provides the highest payoff to the Receiver. This Sender punishment has the Sender being truthful (by Lemma \ref{lem:worstpunishmentsender}) and the Receiver choose a cutoff $x \in (0,1)$ such that 
\[
\rho^S_x(m)=\left\{ \begin{array}{cc} m & \mbox{ if } m < x \\ x & \mbox{ if } m \ge x \end{array} \right.
\]
For any $x$ we can compute Sender and Receiver payoffs. This gives us one condition that any pair of Sender and Receiver in a WRP has to satisfy. We then intersect that condition with the Pareto frontier and obtain an analytic expression for the Receiver maximal WRP payoff on the Pareto frontier. 

\begin{figure}[htb]
    \centering
    \begin{tikzpicture}[scale=0.9]
        \begin{axis}[
            width=14cm,
            height=10cm,
            grid=none,
            %grid style={line width=.1pt, draw=gray!20},
            %major grid style={line width=.2pt, draw=gray!40},
            xlabel={$b$},
            ylabel={Receiver max WRP payoff},
            xmin=0, xmax=5,
            ymin=-0.02, ymax=0, 
            xtick={0,1,2,3,4,5},
            ytick={-0.02,-0.015,-0.01,-0.005,0},
            yticklabel style={
            /pgf/number format/fixed,
            /pgf/number format/precision=3
            },
            scaled y ticks=false,
            legend pos=north east,
            legend cell align={left},
            thick
        ]
            
            \addplot[
                color=black,
                line width=1.5pt
            ] table [x=b, y=uR, col sep=semicolon]{numplot.csv};
               
        \end{axis}
    \end{tikzpicture}
    \caption{The maximal Receiver WRP payoff for $\delta \to 1$ for the \citet{crawford82} game as a function of the bias parameter $b$.}
    \label{fig:WRPmaxCS}
\end{figure}

\begin{figure}[htb]
    \centering
    \begin{tikzpicture}[scale=0.9]
        \begin{axis}[
            width=14cm,
            height=10cm,
            grid=none,
            %grid style={line width=.1pt, draw=gray!20},
            %major grid style={line width=.2pt, draw=gray!40},
            xlabel={$b$},
            ylabel={$\bar{\lambda}$},
            xmin=0, xmax=5,
            ymin=0, ymax=0.2, 
            xtick={0,1,2,3,4,5},
            ytick={0,0.05,0.1,0.15,0.2},
            yticklabel style={
            /pgf/number format/fixed,
            /pgf/number format/precision=2
            },
            scaled y ticks=false,
            legend pos=north east,
            legend cell align={left},
            thick
        ]
            
            \addplot[
                color=black,
                line width=1.5pt
            ] table [x=b, y=L, col sep=semicolon]{numplot.csv};
               
        \end{axis}
    \end{tikzpicture}
    \caption{The weight $\lambda$ on the Sender payoff for the maximal Receiver WRP payoff for $\delta \to 1$ on the Pareto frontier for the \citet{crawford82} game as a function of the bias parameter $b$.}
    \label{fig:WRPmaxCSlambda}
\end{figure}

For $b$ sufficiently small, less than approximately 0.68, one can find a closed-form solution for the Receiver maximal WRP payoff. For $b$ larger than that, one can find it numerically, and one can also analytically show that the Receiver maximal WRP payoff is decreasing for all $b >0$ and that its limit is $-\frac{(1-2^{-2/3})^3}{3} \approx -0.01689$.

Some of the arguments in this proof can be generalized to any full support prior and utility functions determined by a function $f$ with conditions given above. The Receiver punishment works for all such environments. The Sender punishment can also be used, but is not in general the Receiver optimal Sender punishment. This means we do not necessarily find the maximal Receiver WRP payoff on the Pareto frontier in general. We do, however, find a lower bound for the maximal Receiver WRP payoff on the Pareto frontier that is strictly higher than the Sender WRP payoff.   

\subsection{State-independent Sender Preferences}

In this subsection we consider a class of cheap talk games with state-independent Sender preferences.\footnote{Sender state-independent preferences have been considered in static settings in \citet{chakraborty2010persuasion}, \citet{lipnowski2020cheap}, \citet{diehl2021non}, \citet{arieli2023robust}, \citet{steg2023robust}, \citet{whitmeyer2023informedsenderbenefitsreceiver}, \citet{lyu2024informationdesigncheaptalk}, 
and in dynamic settings in \citet{smolin18} and
\citet{jain2026dynamiccheaptalkfeedback}.} Let $\Theta=A=M=\mathbb{R}$.\footnote{We could restrict these sets to be compact subsets of $\mathbb{R}$ to be consistent with the assumptions made in this paper. Doing so would make the analysis a bit more tedious without adding any insights. Moreover, compactness of these sets is not necessary for the results for this application.} Let $u_S(a,\theta)=a$ and $u_R(a,\theta)=-(a-\theta)^2$ and $\theta$ is distributed according some distribution $F$ that is symmetric around zero, has positive variance, and density function $f$. 

In terms of our running story of a government official informing a politician, the incentives of the politician (Receiver) are the same as in the previous (\citet{crawford82}) application, while the official has an even stronger incentive to exaggerate the state: she would like an action as high as possible irrespective of the state. 

The only (Bayes) Nash equilibria of the stage game are babbling equilibria, in which the Sender uses some constant messaging function and the Receiver chooses action zero, the prior mean of the state. This follows from two observations. First, given that the Receiver has a concave utility function, he will always best respond to any belief by choosing a pure action. Second, if the Receiver were to choose more than one pure action in equilibrium, the Sender would prefer one over the other in all states. 

The repeated game exhibits, for sufficiently high discount factors $\delta<1$, the usual folk theorem: at least any payoff profile which gives the Receiver more than his minmax payoff of minus the variance of the state under the prior, can be sustained by the usual subgame perfect Nash reversion (to babbling) grim trigger strategy. In particular, the Receiver obtaining his maximally possible payoff can be sustained as a subgame perfect equilibrium. First we define a notion of dominance between symmetric distributions.

\begin{definition}
Take two symmetric distributions $F$ and $G$ with mean 0, and let $F_+$ and $G_+$ be these distributions truncated below at 0 and normalized by $\frac12$. We say $F$ \emph{dominates} $G$ if $G_+$ dominates $F_+$ in the hazard order.\footnote{A distribution $G_+$ dominates $F_+$ in the hazard order if, given densities $f_+$ and $g_+$, respectively, $\frac{g_+(x)}{1-G_+(x)} \ge \frac{f_+(x)}{1-F_+(x)}$ for all $x$ in the support, with strict inequality for some positive measure of the support. See \citet[Chapter 1.B.1]{shaked2007stochastic} for a thorough discussion.}
\end{definition}

Note that the above definition only applies to symmetric distributions with the same mean, and that it implies the usual second-order stochastic dominance order. That is, if $F$ dominates $G$ in the above sense, then $F$ stochastically dominates $G$ in the second order, or $G$ is a mean-preserving spread of $F$. 

\begin{proposition}\label{prop:stateindep}
Consider the class of games of this section, and any prior distributions $F$ and $G$ over states that are symmetric around zero and have a finite variance. Then there is a value $x_F > 0$ such that the Receiver's supremum WRP payoff for $\delta \to 1$ on the Pareto frontier is $-x_F^2$, the Sender payoff in that WRP equilibrium for $\delta \to 1$ is $x_F$, and the on-path action function in that WRP equilibrium is given by $a(\theta)=\theta + x_F$ for all $\theta$. 

Moreover, if $F$ dominates $G$, then we get that $x_G > x_F$. Thus, the equilibrium action is higher, the Sender payoff is higher, and the Receiver payoff is lower in the Receiver's supremum WRP payoff for $\delta \to 1$ under $G$ than $F$.  
\end{proposition}

The proof is in the appendix. The result implies, for instance, that when $F$ is normally distributed with zero mean and positive variance, an increase in the variance leads to higher payoffs for the Sender and lower payoffs for the Receiver in the Receiver maximal WRP equilibrium. Analogous results also hold for zero-mean $t$-distributions, logistic distributions, and Laplace distributions. For Laplace distributions we can even provide closed-form solutions for all quantities of interest.

To see this, suppose that $F$ is a Laplace distribution centered at zero, i.e., $f(\theta)=\frac{\mu}{2} e^{-\mu |\theta|}$, then the Receiver payoff is $v_R=-\frac{1}{\mu^2} e^{-\mu x}$ and the Receiver maximal WRP condition is given by 
\[
v_R= -\frac{1}{\mu^2} e^{-\mu v_S}.
\]
Combining the PF and WRP conditions (to find the Receiver maximal WRP payoff on the PF) we get 
\[
v_S^2= \frac{1}{\mu^2} e^{-\mu v_S},
\]
which delivers 
\[
v_S(\mu)=\frac{2W(1/2)}{\mu},
\]
and 
\[
v_R(\mu)=-\left(\frac{2W(1/2)}{\mu}\right)^2,
\]
where $W(1/2) \approx 0.35173$ is the Lambert W function evaluated at $\frac12$. 

The comparative statics in $\mu$ (where $\frac{2}{\mu^2}$ is the variance of $\theta$ under the prior $F$) are as follows. The higher $\mu$ (and, thus, the lower the variance) the lower the Sender payoff and the higher the Receiver payoff in the Receiver maximal WRP. 

One can write the Receiver's maximal WRP payoff as $2W(1/2)^2 \approx 0.2475$ times the Receiver's minmax value of $-\frac{2}{\mu^2}$. This means that the value of the Receiver's maximal WRP payoff is about three quarters times the Receiver's overall maximal payoff of $0$ plus a quarter of the value of his minmax payoff. See Figure \ref{fig:stateindep} for the case of $\mu=1$. See Figure \ref{fig:stateindep2} for the player's payoffs in this equilibrium when varying the variance $\frac{2}{\mu^2}$.  

\begin{figure}[htb]
\centering
	\begin{tikzpicture}[scale=0.9]
	\begin{axis}[
		axis lines=middle,
		xlabel={$u_S$},
		ylabel={$u_R$},
		xmin=0, xmax=2,
		ymin=-4, ymax=0.4,
		samples=200,
		domain=0:2,
		width=14cm,
		height=10cm
		]
		
		\addplot[
		black,
		thick
		]
		{
			-x^2
		};

\addplot[
    dotted,
    thick,
    domain=0:2
]
{-2};

\addplot[
    red,
    very thick,
    domain=0.7:{sqrt(2)}
]
{-x^2};

\addplot[
    only marks,
    mark=o,
    mark size=3pt,
    red,
    fill=white
]
coordinates {
    (0.7,-0.49)
    ({sqrt(2)},-2)
};

\node[black] at (axis cs:1.35,-1.2) {WRP};

	\end{axis}

\end{tikzpicture}	
    \caption{The set of payoff profiles of the stage game with state-independent Sender preferences. The minmax value for the Sender is $-\infty$. For the Receiver it is minus one times the variance of $\theta$ under the prior distribution; here for the case of a Laplace distribution centered at zero and scale parameter such that the variance is $2$. The red line segment illustrates the Pareto efficient WRP equilibria for this case.}
    \label{fig:stateindep}
\end{figure}

\begin{figure}[htb]
\centering
\begin{tikzpicture}[scale=0.9]
	\begin{axis}[
		axis lines=middle,
		xlabel={$V$},
		ylabel={$u_R,u_S$},
		xmin=0, xmax=2,
		ymin=-2, ymax=1,
		samples=200,
		domain=0:2,
		width=14cm,
		height=10cm
		]
		
		\addplot[
		black,
		thick
		]
		{
			-0.2474*x
		};

        \addplot[
		dashed,
        black,
		thick
		]
		{
			0.4974*(x^(1/2))
		};

\addplot[
    dotted,
    thick,
    domain=0:2
]
{-x};
	\end{axis}

\end{tikzpicture}	
    \caption{Receiver maximal payoff (solid line) for the state-independent Sender preferences case as a function of the variance $V$ of the state under the assumed Laplace distributed prior. The dotted line is the Receiver's minmax value (as a function of $V$). The dashed line is the Sender payoff in the Receiver maximal WRP payoff case.}
    \label{fig:stateindep2}
\end{figure}

The weight $\lambda \in [0,1]$ on the Sender payoff (to implement payoffs on the Pareto frontier) implicit in the Receiver maximal WRP (and for the Laplace distribution case) is given by $\lambda(\mu)=\frac{4W(1/2)}{\mu +4W(1/2)}$. Note that as $\mu$ goes from zero to infinity (and the variance of the prior goes from infinite to zero), the implied $\lambda$ goes across the whole range from $1$ to $0$. See also Figure \ref{fig:stateindep3}. This means, that high variance cases favor the Sender, low variance cases favor the Receiver in the compromise strategy employed on the equilibrium path of play in the Receiver maximal WRP payoff.

\begin{figure}[htb]
\centering
	\begin{tikzpicture}[scale=0.9]
	\begin{axis}[
		axis lines=middle,
		xlabel={$V$},
		ylabel={$\lambda$},
		xmin=0, xmax=4,
		ymin=0, ymax=1,
		samples=200,
		domain=0:4,
		width=14cm,
		height=10cm
		]
		
		\addplot[
		black,
		thick
		]
		{
			1.407/((2/x)^(1/2)+1.407)
		};

	\end{axis}

\end{tikzpicture}	
    \caption{The weight $\lambda$ on the Sender payoff in the Receiver maximal WRP payoff on the Pareto frontier for the state-independent Sender preferences case as a function of the variance $V$ of the state under the assumed Laplace distributed prior.}
    \label{fig:stateindep3}
\end{figure}

To understand Proposition \ref{prop:stateindep}, an (imperfect) analogy is the standard principal-agent problem with a risk-neutral principal and a risk-averse agent.  There, under a given contract an increase in the volatility of the output can cause incentive constraints to become violated.  If one thinks about the Sender as the (risk-neutral) principal and the Receiver as the (risk-averse) agent, then a similar effect occurs except the incentive constraints are related to renegotiation-proofness.  Moreover, the change in utility goes in the other direction: the Receiver must give up more utility on path to ensure he has an incentive to follow through with punishing the Sender, which becomes more costly as the variance of $\theta$ goes up (or more generally, the distribution changes to one that dominates the original).

Note that the volatility of $\theta$ does not affect the incentives required for an SPE. Renegotiation-proofness constraints are essential to the variance having an impact in Proposition \ref{prop:stateindep}.

\section{Discussion} \label{sec:disc}

\subsection{Relationship to Bayesian Persuasion}

We now explore how WRP payoff profiles (for $\delta \to 1$ and on the Pareto frontier) in a general cheap talk game compare to equilibria of the corresponding Bayesian persuasion game, as studied by \citet{kamenica2011bayesian}. Consider the binary game from Section 4.1, which is essentially the main motivating example of \citet{kamenica2011bayesian}.  As the set of WRP payoffs is not closed, there is no Sender-optimal payoff.  However, the supremum of the WRP payoffs is equal to $2(1-\alpha)$ (for $\alpha \ge \frac12$). This is the Sender payoff obtained at the feasible payoff profile on the Pareto frontier that gives the Receiver his minmax payoff. 

This payoff of $2(1-\alpha)$ is also equal to the payoff the Sender obtains under Bayesian persuasion when the Receiver breaks indifferences in the Sender's favor in this setting.  Indeed, the payoff of the Sender near this point can be thought of as those from the Bayesian persuasion game when the Receiver does not break indifferences fully in the Sender's favor.

This coincidental equality does, however, not generalize to other cheap talk games. In the continuum example from Section 4.2, based on \citet{crawford82}, Bayesian persuasion achieves the Receiver-optimal payoff as demonstrated by \citet[Section V.A.]{kamenica2011bayesian}. By our Theorem \ref{thm:main} this payoff cannot be achieved in a WRP equilibrium of the repeated cheap talk communication game. There does not seem to be a systematic relationship between WRP outcomes of the repeated cheap
talk communication game and equilibria of the one-shot Bayesian persuasion model.

We now explore how WRP equilibria of the repeated cheap talk communication game compare to WRP equilibria of the corresponding repeated Bayesian persuasion game. It suffices to consider again the quadratic \citet{crawford82} example. First note that the repeated Bayesian persuasion game has the same feasible and individually rational set of payoff profiles as the corresponding repeated cheap talk communication game. However, the Receiver-optimal feasible and individually rational payoff profile is, under Bayesian persuasion, an equilibrium of that game. Trivially, this payoff profile is then also WRP, based on the repeated game strategy of playing the stage game equilibrium after all histories. Thus, there does also not seem to be much of a relationship between the WRP payoffs of the two types of repeated games. 

\subsection{Observability of Strategies} \label{sec:observability}

To simplify our analysis we have made strong observability assumptions: we have assumed that players can observe behavioral strategies. In this subsection we argue that not much would change if players could only observe three things: the single realized state, the single realized message that the Sender sent in that state, and the single realized action the Receiver took in that state. 

This would then be a repeated game with imperfect monitoring: players do not observe their opponent's full strategy. Note first, that, given the assumption that states are drawn independently and identically in every period, the repeated game is still such that the game after every history is identical. This means, among other things, that we can still work with the definition of weak renegotiation-proofness given by \citet{maskin89}.\footnote{If there were any persistence in the state, the game would turn from a classical repeated game into a more general stochastic game, and the definition of weak renegotiation-proofness would then have to be adapted.}

Note furthermore that these more stringent observability (or monitoring) assumptions do not change anything about the stage game itself. This means that all statements about the stage game, these include Lemmas \ref{lem:revelation}, \ref{lem:convexF}, \ref{lem:minmax}, \ref{lem:worstpunishmentsender}, and \ref{lem:cocont} continue to hold. The proof of the necessity part of Lemma \ref{lem:characterization} (that generalizes that part of the main theorem of \citet{maskin89} to strategy spaces that are not necessarily compact) still goes through line by line for any monitoring structure that is weaker than the assumption of observable mixtures. Lemmas 1 to 4 and 6, however, are all we need to prove our main result, Theorem \ref{thm:main}. Theorem \ref{thm:main}, therefore, continues to hold under the much weaker monitoring assumptions. 

What is left is the question of whether the sufficiency part of Lemma \ref{lem:characterization}, (i.e. the sufficiency part of the main Theorem in \citet{maskin89}) continues to hold under the weaker monitoring assumption. In Appendix \ref{app:monitoring} we explore this question in detail.  Here we give a summary of our results.

First, consider the set of stage game payoffs attainable in pure strategies, denoted $\mathcal{F}^P$. If the payoffs in the sufficiency part of Lemma 4 are in $\mathcal{F}^P$, and the required strategies are pure strategies, then the original proof goes through directly. That is, if all the sufficient conditions are achieved with pure strategies, then the payoff profile under consideration can be sustained through a WRP equilibrium under the weaker monitoring assumptions. Note that the games from Sections 4.2 and 4.3 have this property.

Second, the case of finite games can be difficult in general. As highlighted by the binary game, $\mathcal{F}^P$ can be quite small compared to $\mathcal{F}$. Hence, randomized (i.e., behavioral) strategies are often required to achieve WRP equilibrium. Therefore, the existence of non-trivial WRP equilibria often does not extend to finite games with sparse feasible pure-strategy payoff sets.

Luckily, we can use the results of \citet{fudenberg91} to establish an approximation result. First, any ex-ante expected payoffs of the repeated game strategy profile can be generated exactly with pure strategies, and all continuation payoffs thereafter on the equilibrium path can be made arbitrarily close. All continuation payoffs can be made to be on the Pareto frontier; they are, therefore, not Pareto rankable.  

For randomized (i.e., behavioral) strategies used in the punishment phase, we run into problems. While one can guarantee that the initial punishment stage is not Pareto rankable with the normal stage payoffs, one cannot guarantee that continuation payoffs along the (finite) punishment path will remain Pareto unrankable. However, we can establish an approximation property we call \emph{$\epsilon$-renegotiation-proof ($\epsilon$-RP)} equilibrium. For an SPE to be $\epsilon$-RP requires that when comparing two continuation equilibria, if the marginal improvement for one agent is greater than $\epsilon$ then the marginal improvement for the other agent is less than $\epsilon$.\footnote{Our notion of $\epsilon$-RP equilibrium is similar in spirit to the notion in \citet{pei25}, though our continuation values need not be near the Pareto frontier.} The concept of an $\epsilon$-RP equilibrium can be motivated through the idea that the two players would face a small $\epsilon>0$ cost of sitting down to renegotiate. Thus, they would renegotiate if both players see a possible payoff improvement of at least $\epsilon$.
The sufficiency part of Lemma \ref{lem:characterization}, then, extends to $\epsilon$-RP equilibrium under our imperfect monitoring assumptions.

\subsection{Implications for Two Player Repeated Games}

While we see this paper as contributing mostly to the literature on strategic communication, the paper also makes some contributions to the literature on repeated games.  First, as mentioned in Remark \ref{rem:necessity2partFM}, our proof of the necessity part of Lemma \ref{lem:characterization} does not depend on the cheap talk structure of the game and does not require compact strategy spaces. We, thus, extend the necessity part of Theorem 1 of \citet{maskin89} to general repeated two-player games with not necessarily finite nor compact strategy sets.

Second, Theorem \ref{thm:main} could be generalized to other two-player games with a continuum of actions (and states if the game has incomplete information) provided three conditions are met: First, there is sufficient continuity in the payoff functions, so that the co-continuity Lemma \ref{lem:cocont} can be established; second, the game has a unique strategy profile that provides a player with their maximal payoff; and third, this strategy profile is not a Nash equilibrium of the stage game. Then, by the same arguments as in our Theorem \ref{thm:main}, this player cannot obtain anywhere close to their maximal payoff in any weakly renegotiation-proof equilibrium of the repeated game (for any discount factor). One should note that \citet{maskin89}, while they do not provide such a result in general, have a series of examples with this feature. 

We believe, however, that the implications of this result in the repeated cheap talk setting make this result especially interesting. Moreover, for the positive results about how the maximal Receiver WRP payoff varies with parameters of the game, we heavily utilize the cheap talk structure, which delivers the crucial Lemmas \ref{lem:revelation} and, especially, \ref{lem:worstpunishmentsender}. These allow us to identify the strategy profiles most suitable for the punishment phase for the Sender, which we can then directly utilize in a series of applications, and could conceivably be utilized in other classes cheap-talk games that we have not explored.

\section{Conclusion} \label{sec:concl}

When can parties effectively communicate in the face of conflicting interests, and to what degree can each benefit from this? In this paper we considered repeated cheap talk between a Sender and a Receiver. We focused on searching for strategies that constitute (weakly) renegotiation-proof equilibria and their associated payoffs. We found that any such equilibrium strategy gives the Receiver a payoff that is bounded away from their optimal (subgame perfect equilibrium payoff). This result holds under fairly weak assumptions, and applies to a wide range of cheap talk environments. Unlike the Receiver's optimal subgame perfect equilibrium payoffs, the Receiver's maximal weakly renegotiation proof payoff then depends on both the Sender's preferences as well as the prior distribution over states. We demonstrate this in three applications, for which we obtain interesting comparative statics of the maximal renegotiation proof Receiver payoffs (and the Sender's payoffs in that equilibrium) when we vary the degree of conflict, either through varying the Sender's preferences or the prior distribution over states.

In terms of information disclosure, for models with large state, action, and message spaces and strictly concave utility functions Pareto efficiency requires full disclosure, whereas in a finite state model it typically need not.  Efficiency can be implemented with truth-telling, but need not: as WRP equilibria require a compromise between the Sender and the Receiver, having the Sender be liberal with the truth and the Receiver believing her (i.e. playing $\beta$) can also be a Pareto efficient WRP equilibrium.

We made several simplifying assumptions in our analysis.  We chose these assumptions with an eye to encouraging cooperation and truth-telling.  We showed that our results essentially go through to the case of imperfect monitoring.  For the assumption of the independence of states across time, we believe that our results are robust (at least in their essence) to more permissive assumptions.  We leave exploring these and other questions for future research.

\appendix

\section{Proofs}

\subsection{Proof of Lemma \ref{lem:characterization}} \label{app:necessity}

\begin{proof}

The original proof of sufficiency is valid here given the convexity of the set of feasible payoffs $\mathcal{F}$ by Lemma \ref{lem:convexF}. The remainder of the proof is for necessity.

Here we prove the following: Let $\delta\in (0,1)$.  A payoff profile $v=(v_S,v_R)$ is WRP equilibrium only if it is feasible and strictly individually rational, $v_S \ge \bar{u}_S, v_R \ge \bar{u}_R$, and there are (mixed) action profiles $(\sigma^S,\rho^S)$ and $(\sigma^R,\rho^R)$ (in the stage game) such that
\[
u_R(\sigma^S,\rho^S) \ge v_R \mbox{ and } \sup_{\sigma' \in B (\mathcal{S})} u_S(\sigma',\rho^S) \le v_S
\]
and
\[
u_S(\sigma^R,\rho^R) \ge v_S \mbox{ and } \sup_{\rho' \in B (\mathcal{R})} u_R(\sigma^R,\rho') \le v_R.
\]

The proof is by contradiction. Fix $\delta \in(0,1)$. Let payoff profile $v=(v_S,v_R)$ be feasible and strictly individually rational, $v_S > \bar{u}_S, v_R > \bar{u}_R$. Suppose there is a repeated game strategy profile $\mathscr{p}$ that generates this payoff profile. Suppose that the necessary conditions are not satisfied. In fact it is sufficient for the proof to consider only one of them.  Let us focus on the first set, which is associated with a Sender punishment. That is, suppose that for all stage game strategy profiles $(\sigma,\rho)$ either:  
\begin{equation} \label{eq:inverseofnecessary}
u_R(\sigma,\rho) < v_R \mbox{ or } \sup_{\sigma' \in B (\mathcal{S})} u_S(\sigma',\rho) > v_S. 
\end{equation}

Let $\Sigma(\mathscr{p})$ denote the set of continuation equilibria of $\mathscr{p}$. Now let 
\[
\underline{v}_S=\inf_{\mathscr{p}' \in \Sigma(\mathscr{p})} u_S(\mathscr{p}').
\] 
There are two cases. Either $\underline{v}_S = v_S$ or $\underline{v}_S < v_S$. As $\mathscr{p} \in \Sigma(\mathscr{p})$ and $u_S(\mathscr{p})=v_S$, we cannot have $\underline{v}_S > v_S$. 

Suppose we have the first case, that is $\underline{v}_S = v_S$. But then at any action profile $(\sigma,\rho)$ that is ever played on the path of play under $\mathscr{p}$ we must have that $u_S(\sigma,\rho)=v_S$. We must also have that there is an action profile $(\sigma,\rho)$ on the path of play under $\mathscr{p}$ at which $u_R(\sigma,\rho) \ge v_R$. But then by Condition (\ref{eq:inverseofnecessary}) we have that $\sup_{\sigma' \in B (\mathcal{S})} u_S(\sigma',\rho) > v_S$. This implies that the Sender has an instantaneous incentive to deviate, which, given that $\mathscr{p}$ is SPE, must be compensated by a continuation equilibrium in $\Sigma(\mathscr{p})$ in which the Sender receives a lower payoff than $v_S$, a contradiction. 

Suppose we have the second case, so that $\underline{v}_S < v_S$. Then for any $\epsilon > 0$, let 
\[
\overline{w}^{\epsilon}_R=\sup_{\mathscr{p}' \in \Sigma(\mathscr{p})} \left\{ u_R(\mathscr{p}') \mid \underline{v}_S \le u_S(\mathscr{p}') \le \underline{v}_S + \epsilon \right\}. 
\]

Let $\epsilon >0$ be sufficiently small so that $\underline{v}_S + \epsilon < v_S$. Then $\overline{w}^{\epsilon}_R \ge v_R$, because for any $\mathscr{p}' \in \Sigma(\mathscr{p})$ with $u_S(\mathscr{p}') \le \underline{v}_S + \epsilon < v_S$ we must have $u_R(\mathscr{p}') \ge v_R$ by $\mathscr{p}$ being WRP. Now, for any $\epsilon,\eta>0$, there must be a $\mathscr{p}''\in \Sigma(\mathscr{p})$ such that $u_S(\mathscr{p}'') \le \underline{v}_S + \epsilon$ and $u_R(\mathscr{p}'') \ge \max\{v_R, \overline{w}^{\epsilon}_R - \eta\}$. Now consider $(\sigma,\rho)$ the stage game strategy profile played at the beginning of $\mathscr{p}''$, that is $(\sigma,\rho)=\mathscr{p}''(\emptyset)$. Suppose first, that the right hand side of Condition \ref{eq:inverseofnecessary} is satisfied for $(\sigma,\rho)$, that is $\sup_{\sigma' \in B (\mathcal{S})} u_S(\sigma',\rho) > v_S$. This gives the Sender an instantaneous incentive to deviate, in which the Sender would gain an additional instantaneous payoff that is at least $v_S-u_S(\mathscr{p}'')> 0$. As $\delta < 1$ and $v_S-u_S(\mathscr{p}'')>0$ (and if $\epsilon$ sufficiently small), for $\mathscr{p}$ to be an SPE the Sender would need to expect a continuation payoff after deviating that is strictly below $\underline{v}_S$, which provides a contradiction.

We, therefore, need to consider that the left hand side of Condition (\ref{eq:inverseofnecessary}) is satisfied, that is $u_R(\sigma,\rho) < v_R$. Recall that $u_R(\mathscr{p}'') \ge \max\{v_R, \overline{w}^{\epsilon}_R - \eta\}$. But then, for sufficiently small $\eta$, the associated continuation payoff (after playing $(\sigma,\rho)$), denote it by $\tilde v_R$, must satisfy $\tilde v_R>\overline{w}^{\epsilon}_R$.  In order to satisfy WRP, we must have that the Sender's continuation payoff in that case, denote it by $\tilde v_S$, satisfies $\tilde v_S \le u(\mathscr{p}'')$. But this contradicts the definition of $\overline{w}^{\epsilon}_R$.

\end{proof}

\subsection{Proof of Theorem \ref{thm:main}}

We first prove a useful Lemma.

\begin{lemma}[Co-continuity at the Top] \label{lem:cocont}
Under Assumption \ref{ass:uniquebest} and that both the Receiver and Sender Bernoulli utility functions are continuous, the Receiver expected utility function, $u_R$, is co-continuous at the (Receiver) top with the Sender expected utility function, $u_S$, by which we mean that for all $\epsilon>0$ there exists an $\eta>0$ such that:
\[
u_R(\sigma,\rho) \ge u_R(\tau,\beta)-\eta \Rightarrow \left\{ \begin{array}{c} u_S(\sigma,\rho) \le u_S(\tau,\beta) +  \epsilon \\ \sup_{\sigma'}u_S(\sigma',\rho) \ge \sup_{\sigma'}u_S(\sigma',\beta) - \epsilon \end{array} \right.
\]
\noindent(where, recall, $\beta$ is the Receiver’s believing strategy and $\tau$ the Sender’s truth-telling strategy).
\end{lemma}

\begin{proof}
Let $\Xi(\mu_0)$ denote the set of joint distributions over $A \times \Theta$ with $\Theta$-marginal $\mu_0$. Given Assumption \ref{ass:uniquebest} and given the Bernoulli utility function $u_R$ is continuous, there is a unique joint distribution $\xi^* \in \Xi(\mu_0)$ that maximizes $\mathbb{E}_{\xi} u_R(a,\theta)$ over all $\xi \in \Xi(\mu_0)$. This joint distribution $\xi^*$ is achieved by the Sender choosing $\tau$ and the Receiver $\beta$, leading to action choices $a^*(\theta)$ for all $\theta \in \Theta$. 

Given the continuity of the expected utility function for the Receiver and the compactness of $\Xi(\mu_0)$ in the weak topology, for any neighborhood $U \subset \Xi(\mu_0)$ of $\xi^*$ there is an $\eta > 0$ such that if $u_R(\sigma,\rho) \ge u_R(\tau,\beta)-\eta$, and, therefore, for the $\xi$ induced by $(\sigma,\rho)$ we have $\mathbb{E}_{\xi} u_R(a,\theta) \ge \mathbb{E}_{\xi^*} u_R(a,\theta)-\eta$, then $\xi \in U$. Given the Bernoulli utility function $u_S$ is also continuous, the expected Sender utility is continuous in the weak topology and we obtain the first inequality.

To establish the second inequality, note first that by Berge's maximum theorem the function $a^*:\Theta \to A$ that maps each state to the Receiver optimal action in that state is continuous, as the Bernoulli utility function $u_R$ is continuous and $A$ and $\Theta$ are compact.\footnote{We cannot use Berge's theorem of the maximum directly to prove the second inequality because the expected Sender utility as a function of the Sender and Receiver stage game strategies is not jointly continuous in the topology of the space of behavioral strategies.} Now fix an $\epsilon > 0$. We will construct some $\eta > 0$ for which the second inequality is true. For the moment consider an arbitrary (but think small) $\eta > 0$. Consider a distribution $\xi \in \Xi(\mu_0)$ over states and actions such that $\mathbb{E}_{\xi} u_R(a,\theta) \ge \mathbb{E}_{\xi^*} u_R(a,\theta)-\eta$. By Lemma \ref{lem:convexF} there is, and by Lemma \ref{lem:worstpunishmentsender} we can restrict attention to, a stage game strategy profile of the form $(\tau,\rho)$ that induces distribution $\xi$. 

For all $a \in A$ and $\theta \in \Theta$ let $z(a,\theta)=u_R(a^*(\theta),\theta)-u(a,\theta)$ denote the Receiver's ``regret'' for playing action $a$ instead of $a^*(\theta)$ in state $\theta$. Note that regret is maximized and equal to zero at $a=a^*(\theta)$, and, therefore, regret is a non-negative function. 

We can derive the expected regret induced by (behavioral) strategy profile $(\tau,\rho)$ for each state by $Z(\theta)=\int_A z(a,\theta) (\rho)_{\theta}(\text{d}a)$ with ex-ante expected regret equal to $\int_{\Theta} Z(\theta)\mu_0(\text{d}\theta) = u_R^* - u_R(\tau,\rho) \le \eta$. 

We can use this regret function to construct a (pure) strategy for the Sender (that is, here, a relabeling of states) that gives the Sender a payoff that is close to her best response payoff given $(\tau,\beta)$. Consider a finite covering of $\Theta$ through open sets $C_i$ with $i \in \{1,...,N\}$ for some finite $N$ and let $\{P_i\}_{i \in \{1,...,N\}}$ denote a partition of $\Theta$, such that each $P_i \subset C_i$. By the assumption that the prior has full support we have that $\mu_0(C_i)>0$ for each $C_i$. We can choose $N$ large to make each $\mu_0(C_i)$ as small as we like. As the regret function is non-negative we have that $\int_{C_i} Z(\theta)\mu_0(\text{d}\theta)$ is small for each $C_i$. This implies that for each $C_i$ there is a state $\theta_i \in C_i$ that produces a small Receiver-regret $Z(\theta_i)$ (for the integral to be small). By choosing $\eta$ small we can increase $N$ and we can make each ``small'' quantity in this paragraph as small as we like. 

By the continuity of $a^*$ and the continuity of the Bernoulli utility $u_R$, we get that $\rho(\theta_i)$ must be close to $\beta(\theta_i)$. This means that, as we choose $\eta$ small we let $N$ get large such that all $\mu_0(C_i)$ get small, for each $\theta_i$ there will be another $\theta_j$ as close to it as we like. 

Let $Q:\Theta \to \Theta$ be such that $Q(\theta)=\theta_i$ if $\theta \in P_i$. Finally, let the sender best response strategy to the Receiver playing $\beta$ be denoted by $\sigma$. Now let $\sigma'$ be derived from $\sigma$ by pushing it through $Q$: for each  state $\theta \in \Theta$, $\sigma'(\theta)$ is given by the pushforward measure of $\sigma(\theta)$ under $Q$. By the continuity of the Bernoulli utility function of the Sender, $u_S$, we get that $u_S(\sigma',\rho)$ is close $u_S(\sigma,\beta)$. As we decrease $\eta$ we can increase $N$ and we can make these two utilities as close as we like. This means we can choose any $\epsilon > 0$, and then choose $N$ accordingly, and then there is a $\eta > 0$ with the desired inequality. 
\end{proof}

\begin{proof}[Proof of Theorem  \ref{thm:main}]

Let $u_S^*=u_S(\tau,\beta)$ and $u_R^*=u_R(\tau,\beta)$ denote the Sender and Receiver's payoffs in the Receiver-optimal strategy profile. Note that $u_R^* \ge u_R(\sigma,\rho)$ for any $\sigma \in B (\mathcal{S})$ and $\rho \in B (\mathcal{R})$. Let $u_S^{**}=\sup_{\sigma'} u_S(\sigma',\beta)$ denote the Sender's optimal deviation payoff when the Receiver uses $\beta$. [Note that when $A^*$ is compact then there is a $\sigma' \in B (\mathcal{S})$ that achieves $u_S^{**}$, but this is not relevant anyway.] By Assumptions \ref{ass:uniquebest} and \ref{ass:minconf} we have that $u_S^{**} > u_S^*$: Assumption \ref{ass:uniquebest} implies that every action $a \in A^*$ is used in the Receiver's strategy $\beta$, and, thus every such $a$ can be induced by the Sender through some message; Assumption \ref{ass:minconf} then implies that there is a set of states that has positive probability under the prior such that the Sender would strictly benefit from inducing some action in $A^*$ that the Receiver would not take in these states. 

Let $\epsilon = \frac13 \left(u_S^{**}-u_S^*\right) >0$. By Lemma \ref{lem:cocont} there is an $\eta>0$ such that for any strategy profile $(\sigma,\rho)$ with $u_R^*-u_R(\sigma,\rho) \le \eta$ we have that $\sup_{\sigma'} u_S(\sigma',\rho) \ge u_S^{**} - \epsilon$ as well as $u_S(\sigma,\rho) \le u_S^* + \epsilon$. We need this argument twice below.

Suppose that there is a WRP payoff profile $(v_S,v_R)$ with $u_R^*-v_R \le \eta$. Then two things must be true. First, by Lemma \ref{lem:convexF}, there is a strategy profile that implements $(v_S,v_R)$ and, thus, by the above argument, 
\[
v_S - u_S^* \le \epsilon.
\]
Then, by Lemma \ref{lem:characterization} there must be a Sender punishment profile $(\sigma^S,\rho^S)$ with $$u_R(\sigma^S,\rho^S) \ge v_R$$ and $\sup_{\sigma'} u_S(\sigma',\rho^S)\le v_S$. As $u_R(\sigma^S,\rho^S) \ge v_R$ we have that $u_R^*-u_R(\sigma^S,\rho^S) < \eta$
and, thus, by the above argument,
\[
u_S^{**} - \sup_{\sigma'} u_S(\sigma',\rho^S) \le \epsilon.
\]
We trivially have that
\[
\sup_{\sigma'} u_S(\sigma',\rho^S)-v_S = \sup_{\sigma'} u_S(\sigma',\rho^S) - u_S^{**} + u_S^{**}-u_S^* + u_S^*-v_S,
\]
and, therefore,
\[
\sup_{\sigma'} u_S(\sigma',\rho^S)-v_S \ge  u_S^{**}-u_S^* - 2 \epsilon,
\]
and, thus,
\[
\sup_{\sigma'} u_S(\sigma',\rho^S)-v_S \ge  \frac13 \left(u_S^{**}-u_S^*\right) >0,
\]
a contradiction. 

\end{proof}

\subsection{Proof of Proposition \ref{prop:binary}}

Let $v=(v_S,v_R)$ be any feasible and individually rational payoff profile. We use the characterization of WRP payoffs in terms of stage game strategies given by Lemma \ref{lem:characterization}. Consider first the punishment for the Receiver $(\sigma^R,\rho^R)$. Let $\sigma^R$ be such that $\sigma^R(0)=\sigma^R(1)=1$ (the Sender always sends message $1$) and $\rho^R=\beta$ (the receiver believes the Sender and chooses action equal to the message received). Then $u_S(\sigma^R,\rho^R)=1$ and $\max_{\rho \in B(\mathcal{R})}u_R(\sigma^R,\rho)=\max\{\alpha,1-\alpha\}$. Thus, $(\sigma^R,\rho^R)$ works as the Receiver punishment for every feasible and strictly individually rational payoff profile $v=(v_S,v_R)$.  

Now consider the punishment for the Sender $(\sigma^S,\rho^S)$. By Lemma \ref{lem:worstpunishmentsender} we can restrict attention to $\sigma^S=\tau$ (i.e. the Sender tells the truth). To determine the Receiver strategy in the Sender punishment, note that any probability weight that we take from the Receiver strategy opposite to always 0 does not change the best-response of the Sender (it is always 1), increases the payoff to the Receiver, and keeps $\max_{\sigma \in B(\mathcal{S})}u_S(\sigma,\rho^S)$ constant. Thus, it is w.l.o.g. that we consider $\rho^S$ as a mixture only between $\beta$ and always playing 0. 

Let $\sigma^S=\tau$ and $\rho^S$ be a mixture of probability $\gamma$ on $\beta$ and $(1-\gamma)$ on always playing 0. We then have $u_R(\sigma^S,\rho^S)=\gamma + (1-\gamma) \alpha$ and $\max_{\sigma \in B(\mathcal{S})}u_S(\sigma,\rho^S)=\gamma$. Now suppose we try to implement an arbitrary payoff profile on the Pareto-frontier, which we can always do by the Sender being truthful and the Receiver putting weight $\nu$ on believe and $(1-\nu)$ on always 1. Then $v_S=\nu (1-\alpha) + (1 - \nu)$ and $v_R=\nu + (1-\nu) (1-\alpha)$. For $(v_S,v_R)$ to be strictly individually rational we need that $\nu + (1-\nu) (1-\alpha) > \alpha$. 

For $(v_S,v_R)$ to be a WRP payoff profile, we, therefore, need that 
\[ \gamma + (1-\gamma) \alpha \ge \nu + (1-\nu) (1-\alpha) \]
and 
\[
\gamma < \nu (1-\alpha) + (1 - \nu)
\]
(Note that the case of a weak inequality never arises in this game.)

We are trying to find for which combination $\nu$ there is a $\gamma \in [0,1]$ such that we can satisfy both inequalities. The second inequality is easier to satisfy the lower we make $\gamma$. The left hand-side of the first inequality is increasing in $\gamma$. Therefore, we get the biggest range of $\nu$ if we make the first inequality an equality. Thus, $\gamma=\frac{1-2\alpha+\alpha \nu}{1-\alpha}$. Plugging this into the second inequality we get $\nu < \frac{1}{2-\alpha}$.

To guarantee that $v_R > \alpha$ (if $\alpha \ge \frac12)$, we also need that $\nu > \frac{2 \alpha-1}{\alpha}$, which is less than $1$ if $\alpha < 1$. Note also that $\frac{2 \alpha-1}{\alpha} < \frac{1}{2-\alpha}$ if $\alpha < 1$. Thus, for $\alpha \ge \frac12$ any Pareto efficient $(v_S,v_R)$ payoff is WRP if and only if $\nu \in (\frac{2 \alpha-1}{\alpha},\frac{1}{2-\alpha})$. If $\alpha < \frac12$ we need to guarantee that $v_R > 1-\alpha$, which is automatically satisfied, and, thus, $\nu \in (0,\frac{1}{2-\alpha})$. Put together, any Pareto efficient equilibrium is, therefore, WRP if and only if the Receiver payoffs are in the range $(\max\{\alpha,1-\alpha\},1-\frac{\alpha(1-\alpha)}{2-\alpha})$. 

\subsection{Proof of Proposition \ref{prop:CS}}

Here we prove a more general statement than Proposition \ref{prop:CS}.

\begin{proposition} \label{prop:CSgen}
Consider the game in Section 4.2 with utility functions $u_R(a,\theta)=f(a-\theta)$ and $u_S(a,\theta)=f(a-\theta-b)$, where $f:\mathbb{R}\to\mathbb{R}$ is a strictly concave, twice differentiable function satisfying $f'(0)=0$ (such that $f$ is maximized at $0$) and with prior belief $\mu_0$ being any distribution over $\Theta$ with a density $g$ that has full support: $g(\theta) > 0$ for all $\theta \in \Theta$. Any WRP payoff profile $(v_S,v_R)$ for $\delta \to 1$ on the Pareto frontier satisfies $v_S\le v_R$. A strategy profile with payoffs on the Pareto-frontier must (except on null sets) induce action $a(\theta)=\theta +\tilde\lambda b$ for all $\theta$ for some $\tilde\lambda \in [0,1]$. If $f$ is symmetric around $0$, then to be WRP for $\delta \to 1$ this strategy profile must have $\tilde\lambda \le \frac12$.
\end{proposition}

\begin{proof}[Proof of Proposition \ref{prop:CSgen}]

Any stage game strategy profile $(\sigma,\rho)$ that implements a payoff profile on the Pareto frontier must induce a distribution over states and actions, for which actions are within $[0,1+b]$ with probability $1$. Construct $\rho'$ from $\rho$ as follows. For any $m \in M$ let $\rho'(m)$ be the action distribution that mimics the action distribution given by $\rho(m)$, but subtracts $b$ from each action.  Then $u_R(\sigma,\rho') \ge u_S(\sigma,\rho)$ by construction.

Now suppose $(v_S,v_R)$ is a WRP payoff profile on the Pareto-frontier. By Lemma \ref{lem:characterization} there must be a (Receiver-punishment) strategy profile $(\sigma^R,\rho^R)$ such that $$u_S(\sigma^R,\rho^R) \ge v_S$$ and $\sup_{\rho'} u_R(\sigma^R,\rho') \le v_R$. But the argument above implies that $\sup_{\rho'} u_R(\sigma^R,\rho') \ge u_S(\sigma^R,\rho^R)$. Therefore, we must have $v_R \ge v_S$. 

We now turn to the equilibrium path of play on the Pareto frontier. A strategy profile $(\sigma,\rho)$ is Pareto-efficient if and only if it maximizes $\lambda u_R(\sigma,\rho) + (1-\lambda) u_S(\sigma,\rho)$ for some $\lambda \in [0,1]$. A Pareto-efficient action function is the function $a(\theta)$ that maximizes
\[
\lambda \int_{\theta} f(a(\theta) - \theta)g(\theta)d\theta + (1-\lambda) \int_{\theta} f(a(\theta) - \theta - b)g(\theta)d\theta.
\]
To maximize the integral, we need to choose $a(\theta)$ to maximize
\[
\lambda f(a(\theta) - \theta) + (1-\lambda) f(a(\theta) - \theta - b)
\]
for almost all $\theta\in\Theta$. 

Noting that any optimal $a(\theta)$ must be in the interval $[\theta,\theta+b]$, the first-order condition is 

\[
 \lambda f'(a(\theta) - \theta) + (1-\lambda) f'(a(\theta) - \theta - b)=0.
\]
Given the strict concavity of the objective function, the solution the first-order condition is a maximum. Consider as a candidate solution $a(\theta)=\theta+\tilde\lambda b$ for some $\tilde\lambda \in [0,1]$. Then the first-order condition becomes 
\[
 \lambda f'(\tilde\lambda b) + (1-\lambda) f'(-(1-\tilde\lambda)b)=0.
\]
Note that this first-order condition does not depend on $\theta$ and any $\tilde\lambda$ that solve it, does so for all $\theta$. For any $\lambda$ the LHS of the first-order condition is non-negative for $\tilde\lambda=0$ (it is $(1-\lambda)f'(-b)$) by $f$ concave with its maximum at $0$, and non-positive for $\tilde\lambda=1$ (it is $\lambda f'(b)$). Thus, by the intermediate value theorem, there is a solution $\tilde{\lambda} \in [0,1]$ that solves the maximization problem.\footnote{Note that if $f(x)=-x^2$, then the unique solution to this problem is $a(\theta)=\theta+\lambda b$, i.e. $\tilde\lambda=\lambda$.}

Finally, the payoff to the Sender and the Receiver for a strategy profile that yields action function $\theta+\tilde\lambda b$ are given by $u_S=f(-(1-\tilde\lambda) b)$ and $u_R=f(\tilde\lambda b)$. If $f$ is symmetric around $0$ then $u_R \ge u_S$ if and only if $\tilde{\lambda} \le \frac12$: $f(-(1-\tilde\lambda) b) = f((1-\tilde\lambda) b) \le f(\tilde\lambda b)$ if and only if $(1-\tilde\lambda) b \ge \tilde\lambda b$ (because $f$ is decreasing for positive arguments). 
\end{proof}

\subsection{Proof of Proposition \ref{prop:CSexist}}

To prove Proposition \ref{prop:CSexist} we here provide two Results, one phrased as a Proposition, the other a Lemma that, together, more than cover it.

\begin{proposition} \label{prop:CSexistgen}
Consider the game in Section \ref{sec:cs} with utility functions $u_R(a,\theta)=f(a-\theta)$ and $u_S(a,\theta)=f(a-\theta-b)$, where $f:\mathbb{R}\to\mathbb{R}$ is a strictly concave, twice differentiable function, symmetric around $0$, satisfying $f'(0)=0$ (such that $f$ is maximized at $0$) and with prior belief $\mu_0$ any distribution over $\Theta$ with a density $g$ that has full support: $g(\theta) > 0$ for all $\theta \in \Theta$. For any bias $b >0$ there is a $\bar{\lambda}\in(0,\frac12)$ such that for any $\lambda \in (\bar{\lambda},\frac12)$ payoff profile $(v_S,v_R)$ with $v_S=f((1-\lambda) b)$ and $v_R=f(\lambda b)$ is WRP for $\delta \to 1$. 
\end{proposition}

\begin{proof}[Proof of Proposition \ref{prop:CSexistgen}]
    
The proof is constructive: we provide Sender and Receiver punishment strategies that satisfy the sufficient condition for WRP as given in Lemma \ref{lem:characterization}.  A note on notation: as all of the strategies in this proof will be pure strategies, and we will use the $\sigma$ and $\rho$ notation to denote pure strategies instead of $s$ and $r$, respectively.

First we consider the Receiver punishment. The Sender uses a partially informative strategy 
\[
\sigma^R_y(\theta)=\left\{ \begin{array}{cc} \mbox{argmax}_{A \in A^*} \mathbb{E}_{\mu_0}[-f(a-\theta)|\theta \in [0,y)] & \mbox{ if } \theta \in [0,y) \\ \theta & \mbox{otherwise} \end{array} \right.
\]
for some $y \in [0,1]$. The Receiver plays the strategy that optimizes the Sender payoff (given the information he has), i.e. $\rho^R_y(m)=m+b$ if $m \in \{\mbox{argmax}_{A \in A^*} \mathbb{E}_{\mu_0}[-f(a-\theta)|\theta \in [0,y)]\} \cup [y,1]$ and $\rho^R_y(m)=0$ otherwise. The Receiver best-response strategy would be to use some $\rho'$ such that $\rho'(m)=m$.

It is easy to see that for all $y \in [0,1]$, $u_R(\sigma^R_y,\rho') = u_S(\sigma^R_y,\rho^R_y)$, consistent with Proposition \ref{prop:CS}. Moreover, as $y$ varies between $0$ and $1$, the Sender payoff $u_S(\sigma^R_y,\rho^R_y)$ varies from the Sender-optimal payoff of $f(0)$ to the Sender-payoff in the babbling equilibrium. 

Thus, for any payoff profile $(v_S,v_R)$ with $v_R > v_S$ with $v_S$ greater that or equal to the Sender-payoff in the babbling equilibrium, there is a $y \in [0,1]$ such that $(\sigma^R_y,\rho^R_y)$ satisfies the part of Lemma \ref{lem:characterization} for the Receiver punishment: we choose $y$ such that $v_R > u_S(\sigma^R_y,\rho^R_y) > v_S$.

Second we consider the Sender punishment. The Sender is truthful, i.e., $\sigma^S=\tau$ (with $\tau(\theta)=\theta)$. The Receiver uses $\rho^{S}_x$ such that $\rho^{S}_x(m)=m$ if $m<x$ and $\rho^{S}_x(m)=x$ if $m \ge x$ for some $x \in [0,1]$. That is, the Receiver punishes the Sender by reverting to a truncation strategy. The Sender best-response is to use strategy $\sigma'(\theta)=\theta+b$.

Note that the strategy profile $(\tau,\rho^S_x)$ induces action function 
\[
a_x(\theta)=\left\{ \begin{array}{cc} \theta & \mbox{ if } \theta < x \\ x & \mbox{ if } \theta \ge x \end{array} \right. \mbox{  } ,
\]
while strategy profile $(\sigma',\rho^{S}_x)$ induces action function
\[
b_x(\theta)=\left\{ \begin{array}{cc} \theta +b & \mbox{ if } \theta < x-b \\ x & \mbox{ if } \theta \ge x-b \end{array} \right.
\]
We then have that $u_R(a_x(\theta),\theta)=f(a_x(\theta)-\theta) \ge f(b_x(\theta)-\theta-b) = u_S(b_x(\theta),\theta)$ for all $\theta$ with strict inequality for all $\theta \ge x-b$. This implies that, as long as $x < 1$, $\max_{\sigma'} u_S(\sigma',\rho^S_x) < u_R(\tau,\rho^S_x)$. 

Note that $u_R(\tau,\rho^S_x)$ ranges from even below the babbling equilibrium payoff for the Receiver for $x=0$ to the Receiver optimal payoff for $x=1$. Now consider payoff profile $(v_S,v_R)$ on the Pareto-frontier with $v_S=f((1-\lambda) b)$ and $v_R=f(\lambda b)$. For $\lambda=\frac12$ there is then an $x \in [0,1]$ such that $u_R(\tau,\rho^S_x)=f(\frac12 b)$ and $ u_S(\sigma',\rho^S_x) < f(\frac12 b)$ for all $\sigma' \in B(\mathcal{S})$. By continuity, there thus also is a $\bar\lambda < \frac12$ such that for any $\lambda \in (\bar{\lambda},\frac12)$ there is an  $x \in [0,1]$ such that $u_R(\tau,\rho^S_x)=f(\lambda b)$ and $\max_{\sigma'} u_S(\sigma',\rho^S_x) < f((1-\lambda) b)$.
\end{proof}

\begin{lemma} \label{lem:senderpunish}
Consider the game from Section 4.2 with quadratic utilities and a uniform distribution. Let $\rho^S_x$ be such that $\rho^S_x(m)=m$ if $m<x$ and $\rho^S_x(m)=x$ if $m \ge x$ for some $x \in [0,1]$. Then, among all behavioral stage game strategy profiles $(\sigma^S,\rho^S)$ with the property that $\sup_{\sigma \in B(\mathcal{S}} u_S(\sigma,\rho^S)=\underline{u}_S$, there is an $x \in [0,1]$ such that $u_R(\tau,\rho^S_x) \ge u_R(\sigma^S,\rho^S)$ for all such $(\sigma^S,\rho^S)$. 
\end{lemma}

\begin{proof}
By Lemma \ref{lem:worstpunishmentsender} we can restrict attention to $\sigma^S=\tau$. To restrict the Sender's supremum payoff, the Receiver must leave out open sets of actions to be played. Suppose, for the moment, that the Receiver chooses one and only one interval of actions within $[0,1]$ of length $\Delta$ that the Receiver does not play (in addition to not playing outside $[0,1]$). Suppose furthermore that, given these constraints, the Receiver plays optimally. If this interval is $[-b,\Delta)$ - let $\rho_{\Delta}$ denote this Receiver strategy - the supremum payoff to the Sender, $\sup_{\sigma \in B(\mathcal{S})} u_S(\sigma,\rho_{\Delta})$ (which is equal to $0$ if $b \ge \Delta$ and equal to $-\frac13 (\Delta -b)^3$ if $b < \Delta$) exceeds the payoff to the Receiver $u_R(\tau,\rho_{\Delta})=-\frac13 \Delta^3$. If the interval is placed in the interior from $(x,x+\Delta)$ such that $x<b$ again the supremum payoff to the Sender exceeds the payoff to the Receiver. If $x>b$ then the Sender supremum payoff is equal to that of the Receiver, both are equal to $-\frac{\Delta^3}{12}$. Finally, if we place this interval at the upper end, i.e., at $(1-\Delta,1+b]$ then the Receiver payoff, of $-\frac13 \Delta^3$, exceeds the Sender supremum payoff, of $-\frac13 (\Delta + b)^3$ (if $b \le 1-\Delta$) and of $-\frac13 (\Delta + b)^3 + \frac13 (b-(1-\Delta))^3$ (if $b>1-\Delta$). All this implies that for any such strategy by the Receiver in which the Receiver plays optimally except for not using an interval of actions of length $\Delta$, we can find a Receiver payoff equivalent Receiver strategy in which the Receiver omits playing an interval of some length $\Delta'$ at the upper end (and does not play outside $[0,1]$), which delivers the worst Sender supremum payoff. 

If there is more than one interval of actions that the Receiver does not play then, analogously, we can find a Receiver payoff equivalent, and at the same time inferior Sender supremum payoff inducing, Receiver strategy that involves omitting an appropriately sized interval of actions at the upper end. From this then follows the statement in the Lemma.
\end{proof}

\begin{proof}[Finishing Proof of Proposition \ref{prop:CSexist}]

By Lemma \ref{lem:senderpunish} we know the optimal Sender punishment strategy profile is from the class of $(\tau,\rho^S_x)$ for some $x \in [0,1]$, with $\rho^S_x(m)=m$ if $m<x$ and $\rho^S_x(m)=x$ if $m \ge x$. We can compute the Receiver payoff $u_R(\tau,\rho^S_x)=-\frac13 (1-x)^3$ and Sender supremum payoff $\sup_{\sigma \in B(\mathcal{S})} u_S(\sigma,\rho^S_x)=-\frac13 (1-x+b)^3$ if $x \ge b$ and $\sup_{\sigma \in B(\mathcal{S})} u_S(\sigma,\rho^S_x)=-\frac13 (1-x+b)^3 + \frac13 (b-x)^3$ if $x<b$.

For the case of $x \ge b$ this translates to any WRP payoff profile $(v_S,v_R)$ having to satisfy 
\[
\sqrt[3]{-3v_S} - \sqrt[3]{-3v_R} = b.
\]
Any WRP on the Pareto frontier satisfies
\[
\sqrt{-v_R}+\sqrt{-v_S}=b.
\]
From the WRP equation we infer that $-v_S \ge -v_R$ or $v_R \ge v_S$. Let $t=\frac{\sqrt{-v_S}-\sqrt{-v_R}}{\sqrt{-v_S}+\sqrt{-v_R}} \in [0,1]$. For a given $t$ we can write $v_S=-\frac{b^2}{4}(1+t)^2$ and $v_R=-\frac{b^2}{4}(1-t)^2$. The WRP equation expressed in terms of $t$ is then given by
\[
\left(\frac{1+t}{2}\right)^{2/3} - \left(\frac{1-t} {2}\right)^{2/3} = \left(\frac{b}{3}\right)^{1/3}.
\] 
The left hand side of this equation is increasing in $t$ and the equation has a unique solution, which is increasing in $b$, provided $b<3$. The implicit solution for the cutoff $x$ satisfies that $x \ge b$ for $b$ between $0$ and approximately $0.6835$. 

For the case of $b>x$ (so roughly $b > 0.6835$), the second part of the Sender supremum expression is relevant. One can then express the system of equations as $x(b)$ being the solution of a polynomial in $x$ (and $b$):
\[
x^6 - 12 b x^4 + (12 b^2 + 6 b) x^3 - 3 b^2 = 0.
\]

To check the limit behavior of the solution of $x(b)$ when $b$ tends to infinity, we can divide the polynomial by $b^2$ to obtain
\[
\frac{x^6}{b^2} - \frac{12}{b} x^4 + \left(12 + \frac{6}{b}\right) x^3 - 3 = 0.
\]
Taking $b \to \infty$ we obtain
\[
12 x^3 - 3 = 0,
\]
which delivers $x=\frac{1}{4^{1/3}}$. Plugging this into the equation for $v_R$ delivers the result for the limiting maximal Receiver WRP payoff.
\end{proof}

\subsection{Proof of Proposition \ref{prop:stateindep}}

\begin{proof}
The Pareto-frontier of this cheap talk game can be found by solving $\max_{a \in \mathbb{R}} \lambda a - (1-\lambda) (a-\theta)^2$ for all $\lambda \in [0,1]$. The first order condition yields $a=\theta + \frac{\lambda}{2(1-\lambda)}$ with $u_S(\lambda)=\frac{\lambda}{2(1-\lambda)}$ and $u_R(\lambda)=-\left( \frac{\lambda}{2(1-\lambda)} \right)^2$. This implies that the Pareto frontier can be written as the set of payoff profiles $(v_S,v_R)$ that satisfy 
\[
v_R=-v_S^2.
\]
Note that the Pareto frontier does not depend on the distribution $F$. 

There is a Receiver punishment strategy, utilizing Lemma \ref{lem:characterization}, that works for all payoff profiles $(v_S,v_R)$ on the Pareto frontier: The Sender tells the Receiver to play $a > v_S$, i.e., $\sigma^R(\theta)=a$ for all $\theta \in \mathbb{R}$ and the Receiver chooses $\rho^R(m)=m$ for all $m$, and, thus plays $a$ for all $\theta$. The maximal payoff the Receiver can obtain from deviating from the punishment is the Receiver's minmax value of minus one times the variance of $\theta$ under the prior, and is thus $ \le v_R$ for all $v_R$ above the Receiver's minmax value. 

To obtain the maximal Receiver WRP payoff we use Lemma \ref{lem:worstpunishmentsender}: w.l.o.g., the Sender punishment has the Sender be truthful, i.e., play $\tau(\theta)=\theta$ for all $\theta \in \mathbb{R}$. For any Receiver strategy $\rho^S \in B(\mathcal{R})$, which can be seen as a function from $\Theta$ to $\mathbb{R}$ (given the Sender is truthful), used to punish the Receiver, as in Lemma \ref{lem:characterization}, the Sender supremum payoff $\sup_{\sigma \in B(\mathcal{S})} u_S(\sigma,\rho^S)$ is simply $\sup_{\theta \in \mathbb{R}} \rho^S(\theta)$. Thus, the best Sender punishment strategy by the Receiver is to play $\rho^S_x(m)$ again, i.e.
\[
\rho^S_x(m)=\left\{ \begin{array}{cc} m & \mbox{ if } m < x \\ x & \mbox{ if } m \ge x \end{array} \right.
\]
Then the maximal payoff to the Sender $v_S=\sup_{\sigma \in B(\mathcal{S})} u_S(\sigma,\rho_x^S)=x$ and the Receiver payoff $u_R(\tau,\rho^S_x)=-\int_{x}^{\infty} (\theta-x)^2 f(\theta) d\theta$. 

The Receiver maximal WRP payoff is obtained for $x$ that solves the system of two equations given by $u_R=-x^2$ (the Pareto frontier condition) and $u_R=-\int_{x}^{\infty} (x-\theta)^2 f(\theta) d\theta$ (the Sender punishment condition). This system can be reduced to solving the equation $x^2=\int_{x}^{\infty} (x-\theta)^2 f(\theta) d\theta$. Note that the function of $x$ that is the right-hand-side of this equation is decreasing in $x$, while the left-hand side is increasing in $x$. At $x=0$ the left-hand side is $0$, while the right-hand side is positive. In the limit when $x \to \infty$ the left hand side tends to $\infty$, while the right hand side tends to zero. Thus, the equation has a unique solution for $x > 0$ proving the first part of the statement of this proposition.

If $G_+$ dominates $F_+$ in the hazard rate order then (using a standard argument, see e.g., \citet[Theorems 1.B.2 and 2.A.1]{shaked2007stochastic}) the right-hand side of the equation $x^2=\int_{x}^{\infty} (x-\theta)^2 g(\theta) d\theta$ (i.e., under $G$) is higher than the right-hand side of $x^2=\int_{x}^{\infty} (x-\theta)^2 f(\theta) d\theta$ (i.e., under $F$) for all $x>0$.\footnote{\citet[Theorem 1.B.2]{shaked2007stochastic} states that the hazard rate order is preserved after applying an increasing function, here the function $(\theta-x)^2$, which is increasing in $\theta$ for all $\theta \ge x$. \citet[Theorem 2.A.1]{shaked2007stochastic} states that if two distributions can be ranked in terms of the hazard rate order, they can also be ranked in the mean residual life order, from which the result follows directly.} This implies that the solution $x_G$ to this equation under $F$ exceeds the solution $x_G$ under $G$, proving the second part of the claim.  
\end{proof}

\section{Imperfect Monitoring}  \label{app:monitoring}

In this Appendix we examine the robustness of our results to our monitoring assumptions.  We define histories as being stage game action-message-state tuples, as opposed to stage game strategies.  As an overview, we have three main takeaways.  First, the necessity part of Lemma 4 extends to weaker monitoring assumptions.  Second, we show that the sufficiency part of Lemma 4 extends to weaker monitoring when the relevant strategies and payoffs are achievable in pure strategies, as is the case with the continuum model in Section 4.2.  

Third, we show that when the payoff space is sparse, such as in the binary game from Section 4.1, then one cannot generally relax our monitoring assumptions.  However, we can achieve an approximation of weak renegotiation-proofness that we call $\epsilon$-weak renegotiation-proofness.  Essentially, any strict Pareto improvements from renegotiation give at least one player an arbitrarily small utility increase.

For our first result, we know that the extreme points of the feasible payoff set $\mathcal{F}$ are achievable in pure strategies.

\begin{lemma} \label{lem:extremepoints}
For cheap talk games the set of feasible (stage game) payoff profiles $\mathcal{F}$ is the convex hull of the set of feasible (stage game) payoff profiles induced by pure strategies. In other words the extreme points of the set $\mathcal{F}$ can be achieved through pure strategy profiles.
\end{lemma}

Recall that players choose strategies before the state is drawn and care about their ex-ante expected payoff in the stage game. In each round the state is redrawn from the same distribution independently of all prior draws (and all prior behavior). Recall that $\pi^t$ denotes the pair of Sender strategy $\sigma^t$ and the composition of the period $t$ (behavioral) strategy profiles $ \rho^t\circ \sigma^t$. We now denote by $\omega^t=(\theta^t,m^t,a^t)$ the tuple of realized state $\theta^t \in \Theta$ drawn according to the prior $\mu_0$, the realized message $m^t \in M$ drawn according to $\sigma^t(\theta^t)$, and the realized action $a^t \in A$ drawn according to $\rho^t(m^t)$. A (restricted)  $t-1$ history of play is now given by $(\omega^0,\omega^1,...,\omega^{t-1})$. Let $\mathcal{H}_r^{t}$ be the set of all possible (restricted) histories up to period $t$, where $\mathcal{H}_r^0=\emptyset$. Let $\mathcal{H}_r=\bigcup_{t=0}^{\infty} \mathcal{H}_r^{t}$ denote the set of all finite (restricted) histories. The (restricted) repeated game strategies are, therefore, functions $\mathscr{s}:\mathcal{H}_r \to B (\mathcal{S})$ and $\mathscr{r}:\mathcal{H}_r \to B (\mathcal{R})$. We continue to denote by $\mathscr{p}=(\mathscr{s},\mathscr{r})$ a (restricted) repeated game strategy \emph{profile}.

The following Lemma states that the same necessary condition identified in \citet{maskin89} for a payoff profile to be WRP is still necessary under the restricted monitoring assumption.

\begin{lemma} \label{lem:characterizationrestrictednecessary}
A necessary condition for a payoff profile $v=(v_S,v_R)$ to be WRP (in the repeated cheap talk game with restricted monitoring for any $\delta < 1$) is that $v$ is feasible and strictly individually rational, $v_S \ge \bar{u}_S, v_R \ge \bar{u}_R$, and there are (behavioral) strategy profiles $(\sigma^S,\rho^S)$ and $(\sigma^R,\rho^R)$ (in the stage game) such that
\[
u_R(\sigma^S,\rho^S) \ge v_R \mbox{ and } u_S(\sigma',\rho^S) \le v_S \mbox{ for all } \sigma' \in B (\mathcal{S})
\]
and
\[
u_S(\sigma^R,\rho^R) \ge v_S \mbox{ and } u_R(\sigma^R,\rho') \le v_R \mbox{ for all } \rho' \in B (\mathcal{R}).
\]
\end{lemma} 

The proof follows from the necessity part of the proof of Lemma \ref{lem:characterization} by noting that monitoring assumptions play no role.

The following Lemma states that the sufficient condition of the original \citet{maskin89} characterization of WRP payoff profiles is still sufficient for WRP payoffs in repeated cheap talk games with the restricted monitoring assumption if the three stage game strategy profiles referred to in the statement of the Lemma, one for implementing the payoff profile in question, the other for the construction of the punishment phases for the two players, are all pure stage game strategy profiles. 

\begin{lemma} \label{lem:characterizationrestrictedpure}
For $\delta < 1$ sufficiently close to $1$ a payoff profile $v=(v_S,v_R)$ is WRP (in the repeated cheap talk game with restricted monitoring) if $v$ is strictly individually rational and there is a pure stage game strategy profile $(\sigma,\rho)$ such that $u_S(\sigma,\rho)=v_S$, $u_R(\sigma,\rho)=v_R$, $v_S > \bar{u}_S, v_R > \bar{u}_R$, and there are pure strategy profiles $(\sigma^S,\rho^S)$ and $(\sigma^R,\rho^R)$ (in the stage game) such that
\[
u_R(\sigma^S,\rho^S) \ge v_R \mbox{ and } u_S(\sigma',\rho^S) < v_S \mbox{ for all } \sigma' \in B (\mathcal{S})
\]
and
\[
u_S(\sigma^R,\rho^R) \ge v_S \mbox{ and } u_R(\sigma^R,\rho') < v_R \mbox{ for all } \rho' \in B (\mathcal{R}).
\]
\end{lemma} 

\begin{proof}[Proof of Lemma 10]

We here follow the approach of \citet{maskin89} in that we construct an appropriate repeated game strategy that has three phases: an on-path ``normal'' phase, and two ``punishment'' phases, one for each player. The constructed repeated game strategy uses exclusively pure stage game strategies.   

Suppose there exist the two punishment stage game strategy profiles $(\sigma^S,\rho^S)$ and $(\sigma^R,\rho^R)$ that satisfy the two conditions. Given these, we construct a repeated game strategy profile that is WRP (or $\epsilon$-WRP) and generates payoffs $(v_S,v_R)$. 

Suppose first that there is a pure stage game strategy profile $(\sigma,\rho)$ with stage game payoff $v$ and that $(\sigma^S,\rho^S)$ and $(\sigma^R,\rho^R)$ are also pure stage game strategy profiles. Then the proof of \citet{maskin89} applies directly: the normal phase is the infinite repetition of $(\sigma,\rho)$, the punishment phase for the Sender is an (appropriately chosen) finite length $T_S$ repetition of $(\sigma^S,\rho^S)$ at the end of which the normal phase resumes, and the punishment phase for the Receiver is an (appropriately chosen) finite length $T_R$ repetition of $(\sigma^R,\rho^R)$ at the end of which the normal phase resumes. Any deviation from any phase by one player leads to the punishment phase of that player. By the same arguments as made by \citet{maskin89} one can show that there is a $\underline{\delta}<1$ such that there are natural numbers $T_R$ and $T_S$ such that this constructed repeated game strategy profile is WRP for every $\delta \in (\underline{\delta},1)$. This proves Lemma \ref{lem:characterizationrestrictedpure}. 
    
\end{proof}

Note that, for instance, all WRP payoffs on the Pareto-frontier for the (generalized) \citet{crawford82} game, identified in Proposition \ref{prop:CSexist}, in the repeated game with the unrestricted monitoring, remain WRP payoffs in the repeated game with restricted monitoring. This follows from the fact that all three stage game strategy profiles used in the proof of Proposition \ref{prop:CSexist} are pure stage game strategy profiles. 

Not all cheap talk games, for instance the binary game of Section \ref{sec:binary}, will have pure stage game strategy profiles that can serve to implement either on path or punishment payoffs. In such games (and, thus, in all cheap talk games) the same conditions of \citet{maskin89} are sufficient, however, for a payoff profile satisfying a slightly weaker notion of renegotiation-proofness. This, we establish in the next lemma. Before we do so, we need to define the slightly weaker notion of renegotiation-proofness. 

\begin{definition}
A subgame perfect equilibrium of a repeated game is $\epsilon$-weakly renegotiation-proof ($\epsilon$-WRP), for $\epsilon > 0$, if for any two continuation equilibria (of that repeated game subgame perfect equilibrium) with payoff profiles $(u^1_S,u^1_R)$ and $(u^2_S,u^2_R)$ we have that $u^1_S > u^2_S + \epsilon$ implies that $u^1_R \le u^2_R + \epsilon$.
\end{definition}
In other words, an $\epsilon$-WRP does not have any two continuation equilibria such that one of these equilibria provides both players with more than $\epsilon$ additional payoff relative to the other equilibrium. There may be Pareto-rankable continuation equilibria, but the payoff difference between them must be small (less than $\epsilon$) for at least one player. One way to think of this definition is that players have a small cost of renegotiating and would, thus, refrain from doing so, if the gains from renegotiation are very small. Note also that when $\epsilon=0$ this definition reduces to the original definition of a WRP. 

The next Lemma says that the sufficiency part of Lemma 4 holds under the new restricted monitoring assumptions if we weaken the definition of renegotiation-proofness from WRP to $\epsilon$-WRP for arbitrarily small $\epsilon > 0$.

\begin{lemma} \label{lem:characterizationrestricted}
For any $\epsilon >0$ there exists $\underline{\delta} < 1$ such that for all $\delta \in (\underline{\delta},1)$ a payoff profile $v=(v_S,v_R)$ is $\epsilon$-WRP (in the repeated cheap talk game with restricted monitoring) if it is feasible and strictly individually rational, $v_S > \bar{u}_S, v_R > \bar{u}_R$, and there are (behavioral) strategy profiles $(\sigma^S,\rho^S)$ and $(\sigma^R,\rho^R)$ (in the stage game) such that
\[
u_R(\sigma^S,\rho^S) \ge v_R \mbox{ and } u_S(\sigma',\rho^S) < v_S \mbox{ for all } \sigma' \in B (\mathcal{S})
\]
and
\[
u_S(\sigma^R,\rho^R) \ge v_S \mbox{ and } u_R(\sigma^R,\rho') < v_R \mbox{ for all } \rho' \in B (\mathcal{R}).
\]
\end{lemma} 

\begin{proof}[Proof of Lemma 11]

To prove Lemma \ref{lem:characterizationrestricted}, we utilize the following version of \citet[Lemma 2]{fudenberg91}.

\begin{lemma}[\citet{fudenberg91}] \label{lem:FM91} For any $\epsilon > 0$ there exists $\underline{\delta} < 1$ such that for all $\delta \in (\underline{\delta},1)$ and every $v \in \mathcal{F}$ there is a deterministic infinite sequence of pure strategies whose discounted average payoffs are $v$, and whose continuation payoffs at each time $t$ are within $\epsilon$ of $v$.
\end{lemma}

The simple idea of the proof of Lemma \ref{lem:characterizationrestricted} is to use the same construction as in Lemma \ref{lem:characterizationrestrictedpure} but replace the stationary randomized (i.e., behavioral) strategy phases with their $\epsilon$ approximation based on Lemma \ref{lem:FM91}. 

Suppose there exist the two punishment stage game strategy profiles $(\sigma^S,\rho^S)$ and $(\sigma^R,\rho^R)$ that satisfy the two conditions. These may now be randomized (i.e., behavioral) stage game strategy profiles. \citet{maskin89} then construct a repeated game strategy profile that has a normal phase and two punishment phases. The normal phase is the infinite repetition of the (behavioral) stage game strategy profile that implements $v$, which we know exists in our cheap talk games by Lemma \ref{lem:convexF}. The punishment phases are length $T_S$ and $T_R$ repetitions of the (behavioral) stage game punishment strategy profiles $(\sigma^S,\rho^S)$ and $(\sigma^R,\rho^R)$, respectively.

By their proof of \citet[Theorem 1]{maskin89} there is a $\underline{\delta}<1$ such that there are natural numbers $T_R$ and $T_S$ such that this constructed repeated game strategy profile is WRP for every $\delta \in (\underline{\delta},1)$.

We here amend this construction so that both players use pure stage game strategy profiles only (so that any (realized) deviation is immediately detected).\footnote{If, say, the sender intends to deviate before knowing the next state $\theta$, but only intends to deviate for some $\theta$'s and none of these $\theta$'s are realized, we here simply see this as no deviation. Such unrealized intended deviations are irrelevant for our construction.} We first pick an arbitrary $\epsilon>0$ and replace the normal phase with the infinite sequence of pure stage game strategy play that exists by Lemma \ref{lem:FM91}. In what follows we make the argument only for the Sender punishment, the Receiver punishment phase can be constructed analogously. For the same $\epsilon$ we then use Lemma \ref{lem:FM91} to construct an infinite sequence of pure stage game strategy profiles such that for all $\delta \in (\underline{\delta},1)$ all continuation payoffs of that sequence are within $\epsilon$ of $u_S(\sigma^S,\rho^S)$ and $u_R(\sigma^S,\rho^S)$.\footnote{Technically this $\underline{\delta}$ could be different from the one used in the \citet{maskin89} construction, but (given the sequence of quantifiers) we can just choose the maximum of the two.} Let $\{u_t\}_{t=0}^{\infty}$ be the infinite sequence of Sender stage game payoffs in this construction. The Sender punishment phase is then given by the first $T_S$ rounds of this infinite sequence, after which play resumes with the normal phase.

What remains to be shown is that all continuation payoffs along the Sender punishment phase are within an $\epsilon'$ (that can be made arbitrarily small when we choose $\epsilon$ arbitrarily small) of the punishment phase using the same behavioral stage game strategy profile $(\sigma^S,\rho^S)$ in the \citet{maskin89} construction. 

By construction we have that $(1-\delta)\sum_{t=t'}^{\infty} \delta^{t-t'}u_t$ is within $\epsilon$ of $u_S=u_S(\sigma^S,\rho^S)$ for all $t'$. 

We need to compare $(1-\delta)\sum_{t=t'}^{T_S} \delta^{t-t'}u_t + \delta^{T_S-t'+1} v_S$ and $(1-\delta)\sum_{t=t'}^{T_S} \delta^{t-t'} u_S + \delta^{T_S-t'+1} v_S$ for any $t' \le T_S$. We have 

\begin{align*}
& (1-\delta)\sum_{t=t'}^{T_S} \delta^{t-t'}u_t + \delta^{T_S+1-t'} v_S  - (1-\delta)\sum_{t=t'}^{T_S} \delta^{t-t'} u_S - \delta^{T_S-t'+1} v_S & \\
= & (1-\delta)\sum_{t=t'}^{T_S} \delta^{t-t'}u_t  - (1-\delta)\sum_{t=t'}^{T_S} \delta^{t-t'} u_S & \\
= & (1-\delta)\sum_{t=t'}^{\infty} \delta^{t-t'}u_t - (1-\delta)\sum_{t=T_S+1}^{\infty} \delta^{t-t'}u_t
- (1-\delta)\sum_{t=t'}^{\infty} \delta^{t-t'} u_S + (1-\delta)\sum_{t=T_S+1}^{\infty} \delta^{t-t'} u_S & \\
 = & (1-\delta)\sum_{t=t'}^{\infty} \delta^{t-t'}u_t - \delta^{T_S+1-t'} (1-\delta)\sum_{t=T_S+1}^{\infty} \delta^{t-(T_S+1)} u_t - u_S + \delta^{T_S+1-t'} u_S & \\
= & \left((1-\delta)\sum_{t=t'}^{\infty} \delta^{t-t'}u_t - u_S \right) + \delta^{T_S+1-t'} \left( u_S - (1-\delta)\sum_{t=T_S+1}^{\infty} \delta^{t-(T_S+1)} u_t \right). &
\end{align*}

By construction we have that for any $t' \le T_S$, $|(1-\delta)\sum_{t=t'}^{\infty} \delta^{t-t'}u_t - u_S| < \epsilon$ and also $|(1-\delta)\sum_{t=T_S+1}^{\infty} \delta^{t-(T_S+1)} u_t - u_S| < \epsilon$. By the triangle inequality we then have that $|(1-\delta)\sum_{t=t'}^{T_S} \delta^{t-t'}u_t - (1-\delta)\sum_{t=t'}^{T_S} \delta^{t-t'} u_S| < \epsilon (1 + \delta^{T_S+1-t'}) < 2 \epsilon$.  Thus, for any $\epsilon > 0$ the so constructed repeated game strategy profile is $4\epsilon$-WRP. 

We finish the proof by noting that these strategies (under imperfect monitoring) constitute subgame perfect Nash equilibria.  This follows from the fact that all relevant inequalities that define equilibrium under observable behavioral strategies are strict.  Hence, achieving nearby payoffs in all cases remains a SPE.\end{proof}

\printbibliography

\end{document}